\documentclass[11pt]{article}

\usepackage[margin=1in]{geometry}
\usepackage{setspace}
\usepackage[font=small]{caption}
\usepackage{amsmath,amssymb}
\usepackage{booktabs}
\usepackage{array}
\usepackage{graphicx}
\usepackage{microtype}
\usepackage{placeins}
\usepackage{flafter}

\usepackage[authoryear,round]{natbib}
\PassOptionsToPackage{hyphens}{url}
\usepackage[colorlinks=true,linkcolor=blue,citecolor=blue,urlcolor=blue]{hyperref}

\graphicspath{{fig/}}
\newcommand{\incfig}[2][]{\includegraphics[#1]{#2}}

\DeclareMathOperator{\tr}{tr}
\newcommand{\sph}{S_{-1}}
\newcommand{\Run}{\mathbf{Q}}
\newcommand{\Usupp}{\mathcal{U}}

\title{A Sphericity Filter for Space-Filling Designs in Nondeterministic Experiments}
\author{Andrew T. Karl\thanks{Karl Statistical Services LLC, Aurora, CO, USA. Corresponding author. Email: \href{mailto:akarl@asu.edu}{akarl@asu.edu}. ORCID: \href{https://orcid.org/0000-0002-5933-8706}{0000-0002-5933-8706}.}}
\date{}

\begin{document}
\maketitle

\begin{abstract}
\noindent
Space-filling designs support flexible modeling in noisy multi-response experiments, but an individual design can give poor precision for low-order effects. We use a low-order working model to identify effects whose estimation precision we seek to protect, while allowing a more complex final response model. The filter selects one complete design from a candidate pool by minimizing a geometric loss divided by sphericity, without exchanging or altering its points. Sphericity measures the balance of the working-model confidence ellipsoid's principal-axis variances, independently of its volume. In a working-model basis orthonormal over the prediction region, sphericity is the shape factor of the I-criterion, which measures average prediction variance. In cube studies with quadratic least-squares and Gaussian-process fits, this target filter gives modest average prediction gains over the geometrically best candidate at small geometric cost. Gains in working-model precision are largest near saturation: in a four-factor example, worst-decile prediction standard error falls by about a quarter relative to the geometrically best candidate. In constrained mixtures, raw coefficient-based filtering and A-optimality can emphasize extrapolated Scheff\'e coefficients and yield poor regional prediction precision. Target filtering improves regional working-model precision in most settings studied, mirroring the advantage of I- over A-optimality for highly constrained mixture designs.

\end{abstract}


\section{Introduction}\label{sec:intro}

Space-filling designs support flexible modeling of complex responses \citep{McKayBeckmanConover1979,JosephGulBa2015} in computer experiments and, increasingly, in noisy physical experiments and stochastic simulations \citep{bridge,ubridge}. For example, they often initialize Bayesian optimization \citep{HvarfnerEtAl2025} and related sequential workflows \citep{KarlSVEMnet2026}, where the initial design can affect subsequent performance. In multi-response experiments, some responses may follow simple relationships even when others require flexible models. Individual space-filling designs can, by chance, give poor precision for jointly estimating plausible low-order effects. We screen complete designs using a working low-order linear model containing main effects and, where appropriate, two-factor interactions or quadratic terms.

The motivation for this filter arose from pharmaceutical formulation studies that began with fast flexible filling (FFF) designs. \citet{KarlEtAlJoVE2023} recommend FFF as an accessible starting point for mixture--process studies, emphasizing construction ease, response-model flexibility and interior coverage. Interior coverage helps avoid overcommitting runs to boundary formulations that may fail when formulation feasibility and response magnitude are being explored together. Follow-up runs were chosen by Gaussian-process (GP) Bayesian optimization in JMP Pro or SVEMnet's ensemble-based optimization \citep{KarlSVEMnet2026}. All assays were analyzed flexibly with GP or self-validated ensemble models (SVEM) \citep{LemkusEtAl2021}. The observed responses ranged from nonlinear drop-offs to clear main-effect and two-factor-interaction structure. In some FFF designs, unexpectedly strong correlations among low-order terms left effects such as a buffer-type-by-pH interaction poorly supported. The proposed filter automates selection of a space-filling design that supports low-order effects while making small observed concessions in space-filling geometry.

Choosing among complete space-filling candidates is established practice for random Latin hypercubes \citep{HernandezLucasSanchez2012} and mixture designs \citep{BorkowskiPiepel2009}. Software-generated designs can vary substantially in quality \citep{LucasParker2023}. \citet{PiepelEtAl2019}, for example, compared 20 FFF designs for a constrained mixture region on geometric criteria before making a final choice.

Other approaches change the construction itself, targeting nonuniform coverage \citep{YangLuAndersonCook2025} or combining space filling with model-based criteria. Orthogonal-maximin Latin hypercubes balance correlation and separation criteria \citep{JosephHung2008}, while bridge and unrestricted bridge (U-bridge) designs trade space filling against D- or I-optimality for a specified working model \citep{bridge,ubridge}. I-optimal runs can also augment a space-filling design after initial responses suggest that a polynomial is adequate \citep{JohnsonEtAl2010}. Our filter selects a complete candidate from a pool using geometry and the shape of its working-model confidence ellipsoid. It leaves the generated design points unchanged. The working model specifies effects whose joint estimation we seek to protect, without restricting the final response analysis to that model.

The shape term comes from factorizations of the A- and I-criteria. \citet{JonesAllenMoyerGoos2020} showed that similar D-efficiency can conceal differences in coefficient variances and power, motivating A-optimality for screening \citep[see also][]{StallrichAllenMoyerJones2023}. \citet{KarlSphericity2026} expresses A-efficiency as the product of D-efficiency and sphericity, separating determinant scale from the balance of principal-axis variances. Raw sphericity, the A-criterion's shape factor, uses fixed coefficient coordinates. Target-relative sphericity, the I-criterion's shape factor, uses coordinates matched to a prediction region.

In constrained mixtures, model coding requires particular care. In the Scheff\'e parameterization \citep{Scheffe1958}, linear coefficients describe pure-component responses that can require extrapolation far outside the feasible region \citep{Piepel2006}, even though every design run lies within it. A-optimal mixture designs depend on how the model is scaled \citep{DraperPukelsheim1998}. I-optimal mixture designs target prediction over the region instead \citep{GoosJonesSyafitri2016,CoetzerHaines2017}. We show how this distinction between coefficient precision and regional prediction carries through to raw and target-relative filtering, and how their performance in constrained mixtures mirrors the corresponding A- and I-optimal designs.

\section{The sphericity filter}\label{sec:background}\label{sec:filters}\label{sec:rule}

We generate a pool of $K$ complete space-filling designs by repeating the same stochastic construction with the run size, region and construction settings held fixed.

\subsection{Geometric loss and candidate designs}\label{sec:sfd}\label{sec:ties}

Let $\Run$ be a candidate's $n\times k$ run matrix, with one row $x^\top$ for each run. Its unique support is $\Usupp(\Run)=\{u_1,\ldots,u_{n_U}\}$, with $n_U$ distinct settings. We write $g(\Run)>0$ for a geometric loss evaluated on this unique support, on a fixed ratio scale with smaller values favorable. Replicates do not change the unique support. Best geometry selects the candidate with the smallest $g$. We use two losses, depending on the region and factor types.

FFF clusters a large random sample from the region, with one cluster per run \citep{FFF,LekivetzJones2019}. JMP supplies each cluster's centroid or a point chosen using maximum projection (MaxPro), its default \citep{JMPFFFDesignDetails19}. MaxPro thus helps construct candidates and supplies our cube selection loss. With factors scaled to $[0,1]$ for geometry, continuous MaxPro is \citep{JosephGulBa2015}
\begin{equation}\label{eq:maxpro}
\mathrm{MaxPro}(\Usupp(\Run))=
\left\{\binom{n_U}{2}^{-1}\sum_{i<j}
\prod_{l=1}^{k}(u_{il}-u_{jl})^{-2}\right\}^{1/k}.
\end{equation}
It emphasizes projections and requires $n_U\ge2$ with no shared coordinate levels. Removing duplicate rows does not remove all ties. Rounding and discrete factors create them in the lipid nanoparticle (LNP) application.

For mixtures we select on regional coverage instead. For a regional reference sample $R=\{r_1,\ldots,r_M\}$ shared across candidates, average coverage distance is
\begin{equation}\label{eq:acd}
\mathrm{ACD}(\Usupp(\Run))=M^{-1}\sum_{m=1}^{M}
\min_{u\in\Usupp(\Run)}\lVert r_m-u\rVert.
\end{equation}
ACD averages the distance from each regional reference point to the nearest design point \citep{FFF}. \citet{BorkowskiPiepel2009} use its range-scaled form for mixtures.

\subsection{Working model and raw sphericity}\label{sec:criteria}

Good space filling can coexist with weak support for particular response relationships. A working model approximates the mean response by $f(x)^\top\beta$, where the fixed regression vector $f(x)\in\mathbb{R}^p$ contains the effects to be protected. The factorization below applies to any such basis with a full-column-rank model matrix, including interactions, powers and fixed spline terms.

For a candidate $\Run$, the $n\times p$ working-model matrix $X$ has one row $f(x)^\top$ for each run $x$. Thus $\Run$ records the experimental settings, while $X$ evaluates the chosen model terms at those settings. The factor coding used to form $X$ need not match the geometry scaling: our cube model uses $[-1,1]$, while MaxPro uses $[0,1]$. We evaluate candidates using the same working model, coding and geometric loss, discarding any candidate whose working-model matrix $X$ is not of full column rank. Full column rank, $\operatorname{rank}(X)=p$, requires $n\ge p$ and at least $p$ distinct settings. The pool size $K$ counts eligible candidates.

The information matrix $C=C(\Run)=X^\top X$ can be computed before responses are observed, and all runs, including replicates, contribute to it. Under independent errors with common variance $\sigma^2$, the least-squares coefficient covariance gives $\Sigma=C^{-1}=\operatorname{Var}(\hat\beta)/\sigma^2$. The D- and A-criteria \citep[Sec.~6.1]{Pukelsheim2006} are
\begin{equation}\label{eq:defs}
D=\det(C)^{1/p},\qquad A=\tr(C^{-1}).
\end{equation}
Larger $D$ and smaller $A$ are favorable. JMP Evaluate Design reports $D_{\mathrm{eff}}=100D/n$ and $A_{\mathrm{eff}}=100p/(nA)$ \citep{JMPDesignEfficiency2025}, both 100 at $C=nI_p$. Following \citet{KarlSphericity2026}, define raw sphericity
\begin{equation}\label{eq:fact}
S_{-1}(C)=\frac{p}{A D}=\frac{A_{\mathrm{eff}}}{D_{\mathrm{eff}}}\in(0,1].
\end{equation}
Equivalently, $A/p=D^{-1}/S_{-1}(C)$. Sphericity is the geometric-to-arithmetic mean ratio of the eigenvalues of $\Sigma$, with the same functional form as the $p$th root of Mauchly's statistic: $S_{-1}=W^{1/p}$ when $W$ is evaluated on $\Sigma$ \citep{Mauchly1940}. It equals one when the coefficient confidence ellipsoid is spherical and is unchanged by common information scaling. It measures shape, not size: sphericity alone need not give small variances or good coverage. Raw sphericity weights Euclidean coefficient coordinates equally and generally changes under recoding \citep[Sec.~3.3]{KarlSphericity2026}.

\subsection{Prediction targets and target-relative sphericity}\label{sec:target}

When the aim is prediction over a region, we can measure uncertainty balance in coordinates matched to that target. Let $\nu$ be a probability measure describing where predictions matter. Its moment matrix depends on the working model and target, not the candidate:
\begin{equation}\label{eq:mixturetarget}
B=\int f(x)f(x)^\top\,d\nu(x),\qquad I_\nu=\tr(C^{-1}B).
\end{equation}
The I-criterion averages working-model prediction variance, omitting $\sigma^2$, so smaller values are favorable \citep{MyersMontgomeryAndersonCook2009,JonesAllenMoyerGoos2020}. For finite positive-definite $B$ and positive-definite $C$, factor $B=LL^\top$ and define the target-coordinate information matrix $C_B=L^{-1}CL^{-\top}$. In these coordinates, the A-criterion equals $I_\nu$ \citep[Sec.~5.3]{FilovaHarman2020} and the D-criterion is $D_\nu=(\det C/\det B)^{1/p}$. The target-relative factorization of \citet[Sec.~4.4]{KarlSphericity2026} gives
\begin{equation}\label{eq:targetfactor}
S_{-1}(C_B)=\frac{p}{I_\nu D_\nu}\in(0,1].
\end{equation}
It equals one exactly when $C$ is proportional to $B$. A nonsingular reparameterization changes $C$ and $B$ together and leaves $I_\nu$, $D_\nu$ and $S_{-1}(C_B)$ unchanged \citep[Sec.~4.4]{KarlSphericity2026}. Using target coordinates aligns the precision target with regional prediction. For cubes, $\nu$ is uniform on $[-1,1]^k$ and we write $I$ for $I_\nu$. Mixture and joint LNP targets are specified in Section~\ref{sec:mixtureapplication}.

\subsection{Combining geometry and sphericity}\label{sec:compute}

With the geometric loss and prediction target fixed, we select the candidate minimizing
\begin{equation}\label{eq:score}
\mathrm{score}(\Run)=\frac{g(\Run)}{\sph(C_B(\Run))}.
\end{equation}
This target filter ranks the entire candidate pool with one score. Since $0<S_{-1}(C_B)\le1$, its reciprocal is a multiplicative penalty for imbalance, equal to one for a spherical ellipsoid. A proportional gain in sphericity offsets the same proportional increase in geometric loss. This follows the structure of the A- and I-factorizations: geometric loss takes the place of inverse determinant scale, with the same sphericity penalty.

$B$ can be obtained from analytic regional moments \citep[Sec.~3.2]{GoosJonesSyafitri2016} or approximated by a weighted average of $f(x)f(x)^\top$ over sites representing the prediction target. Equations~\eqref{eq:targetfactor} and~\eqref{eq:score} give $g/S_{-1}(C_B)=gI_\nu D_\nu/p$. Since $D_\nu$ differs from $D$ by a pool-wide constant, ranking by $gI_\nu D$ gives the same selection. This is a ranking equivalence: calculating $I_\nu$ still requires the target moments. The supplement gives JMP coding and moment-normalization instructions.

The filter screens for uncertainty imbalance without rewarding uniform precision gains. A secondary control, the $gI$ selector, minimizes geometry times the complete $I_\nu$. Because $I_\nu=p/\{D_\nu S_{-1}(C_B)\}$, it also rewards determinant scale, placing greater emphasis on the complete working-model prediction criterion. We compare the two to assess the resulting precision gains and geometric costs. At equal geometry and I, the target filter prefers smaller determinant scale and hence more balanced variances.

The raw filter uses the A-criterion's shape factor to balance uncertainty in the working-model coefficients \citep{KarlSphericity2026}. It gives $g/S_{-1}(C)=gAD/p$ and needs no regional moment matrix. In the tested cube coding it gave slightly better working-model precision than target filtering at about twice the median MaxPro cost. Its ranking generally changes under recoding, so these results do not make it a regional-prediction shortcut for constrained regions.

In practice, applying the filter takes four steps.
\begin{enumerate}
\item Choose the region, run size, working-model basis and coding, prediction target, geometric loss and pool size $K$, the candidate budget. Our main comparisons use 500 candidates per pool.
\item Generate complete candidates and retain those with full-column-rank $X$. Use all runs for information and unique support for geometry, with a tie-safe loss such as ACD when factor levels repeat.
\item Compute the score in Equation~\eqref{eq:score} for every candidate and select its minimum over the full pool.
\item Report the percentage changes in geometric loss and working-model prediction variance relative to best geometry.
\end{enumerate}

\section{Illustration with an FFF pool}\label{sec:fffpool}

\begin{figure}[!htbp]
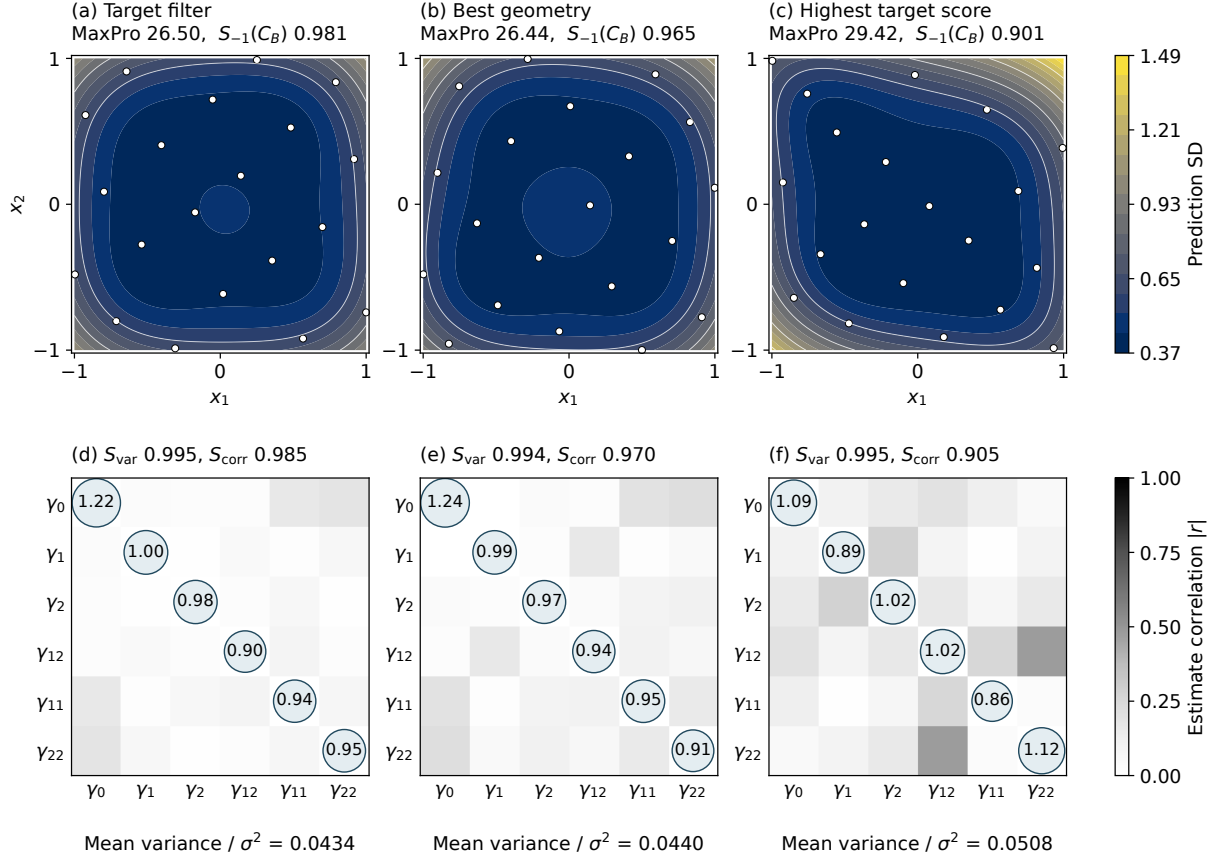

\centering
\incfig[width=0.98\linewidth]{confirmation/fff_pool_229_comparison.pdf}
\caption{Three candidates from one 500-candidate FFF pool. (a--c) Prediction standard deviation with unit error variance, a shared scale and the runs overlaid. (d--f) Target-coordinate estimate correlations (tile shade gives magnitude; signs omitted) and relative variances (circle area and number). Mean target-coordinate variance, $I/6$, appears below each map.}
\label{fig:fff}
\end{figure}

We searched the 500 two-factor, 20-run FFF pools used in Section~\ref{sec:sim} for an example of unusually poor working-model support among the generated candidates. Figure~\ref{fig:fff} compares the target-filter selection, best geometry and the candidate with the highest target score in this pool. The working model is full quadratic on $[-1,1]^2$, with $f=(1,x_1,x_2,x_1x_2,x_1^2,x_2^2)^\top$ and a uniform prediction target.

Prediction standard deviation is $\sqrt{f(x)^\top C^{-1}f(x)}$. The coefficient maps use $f(x)^\top L^{-\top}$, a region-orthonormal basis obtained from the Cholesky factorization $B=LL^\top$. For the diagonal $d$ and signed correlation matrix $R$ of $C_B^{-1}$, the maps show relative variances $d_j/\operatorname{AM}(d)$ and the components $S_{\mathrm{var}}=\operatorname{GM}(d)/\operatorname{AM}(d)$ and $S_{\mathrm{corr}}=\det(R)^{1/6}$. The component split depends on these axes. Total target sphericity does not. The mean variance below each map gives the overall uncertainty scale, with smaller values indicating greater average precision.

The highest-score candidate has elevated prediction uncertainty near two opposite corners and stronger correlations among coefficient estimates. Its marginal-variance balance is similar to the other two designs, but its correlation component is appreciably worse. Both selectors avoid this candidate. In this pool, the target filter reduces average working-model prediction variance by 1.23\% relative to best geometry at a 0.24\% increase in MaxPro (Table~\ref{tab:fff}). The repeated-pool studies below assess the precision gains and geometric costs of filtering, including the risk of selecting designs with unusually poor working-model precision.

\begin{table}[!htbp]
\centering\small
\setlength{\tabcolsep}{4pt}
\begin{tabular}{lcccc}
\toprule
Design & $\sph(C_B)$ & MaxPro & MaxPro$/\sph(C_B)$ & $I_{\mathrm{eff}}$ \\
\midrule
Target filter & 0.981 & 26.50 & 27.03 & 92.3 \\
Best geometry & 0.965 & 26.44 & 27.40 & 91.2 \\
Highest (worst) target score & 0.901 & 29.42 & 32.65 & 79.0 \\
\bottomrule
\end{tabular}
\caption{Diagnostics for the three designs in Figure~\ref{fig:fff}. Smaller MaxPro and target score are favorable. Larger sphericity and $I_{\mathrm{eff}}$ are favorable. Here $I_{\mathrm{eff}}=100\min_{d\in\mathrm{pool}}I(d)/I$ is relative to the best I in this pool.}
\label{tab:fff}
\end{table}

\section{Prediction and working-model precision}\label{sec:sim}

Table~\ref{tab:studymap} organizes the simulation studies presented in the paper. Alongside average prediction risk, we examine precision tails: the least precise designs selected across repeated pools.

\begin{table}[!htbp]
\centering\small
\renewcommand{\arraystretch}{1.12}
\begin{tabular}{@{}>{\raggedright\arraybackslash}p{.22\linewidth}>{\raggedright\arraybackslash}p{.53\linewidth}>{\raggedright\arraybackslash}p{.18\linewidth}@{}}
\toprule
Question & Evidence & Results \\
\midrule
Average prediction and precision tails & Common pools: 500 per setting in six cubes for OLS, GP and working-model-tail comparisons. Six bounded-mixture settings also have 500 pools. & Table~\ref{tab:responseresults}; Figures~\ref{fig:risksummary}--\ref{fig:centertail}. \\
Scope of the protection & Three other generators and four larger FFF cube settings, each with 500 pools per setting. & Section~\ref{sec:additionaldesigns}; Table~S5. \\
Choice of coordinates & Matched direct A/I references for all six cubes, six bounded mixtures and fresh LNP, with 500 pools per setting. & Section~\ref{sec:mixtureapplication}; Figure~\ref{fig:coordinates}. \\
Diagnostic reference & Reference cohort: 50 pools per cube setting, 30 per bounded mixture and 30 LNP; fixes tail scale $m$ and supplies Pareto and scale--shape diagnostics. & Figure~\ref{fig:candidatesphericity}; Table~S4; Section~S2. \\
\bottomrule
\end{tabular}
\caption{Summary of the simulation studies.}
\label{tab:studymap}
\end{table}

\subsection{Study design}\label{sec:simdesigns}

The cube response comparisons and working-model precision tails use the same complete FFF pools. We use the six settings of \citet{ubridge}: $n=10,20$ for $k=2$, $n=15,30$ for $k=3$, and $n=20,30$ for $k=4$. Each setting has 500 pools of 500 candidates, generated in JMP Pro 19. Each factor is coded on the fixed interval $[-1,1]$, irrespective of its realized range. The working model is a full quadratic response surface model in these coded factors.

The procedures in Table~\ref{tab:procedures} were applied to every pool. Each pool selector is compared with best geometry (the primary comparator) and with a single unscreened FFF draw. For OLS, single-draw risk averages all candidates in a pool. For GP, it uses the first accepted candidate. Averaging all candidates makes the OLS estimate more precise.

\begin{table}[!htbp]
\centering\small
\begin{tabular}{@{}>{\raggedright\arraybackslash}p{.29\linewidth}>{\raggedright\arraybackslash}p{.65\linewidth}@{}}
\toprule
Procedure & Criterion minimized \\
\midrule
\multicolumn{2}{@{}l}{\textbf{Select from a common pool of complete candidates}} \\[2pt]
Single FFF draw & None: no screening \\
Best geometry & $g$ \\
Target filter & $gI_\nu D_\nu\propto g/S_{-1}(C_B)$ \\
Raw filter & $gAD\propto g/S_{-1}(C)$ \\
$gI$ selector (control) & $gI_\nu$ \\
\midrule
\multicolumn{2}{@{}l}{\textbf{Construct a reference separately for each pool}} \\[2pt]
I-U bridge & $I_\nu$; runs chosen from a $20n$-point MaxPro FFF support (cube study) \\
Direct I-optimal & $I_\nu$ over the feasible design space \\
Direct A-optimal & $A$ over the feasible design space \\
\bottomrule
\end{tabular}
\caption{Procedures compared. Pool products are ranking-equivalent within a fixed model, coding and prediction target. Here $g$ is MaxPro on cubes and ACD in mixtures. Supplementary Section~S1 adds the A-U bridge and OLS-only minimum-I pool candidate.}
\label{tab:procedures}
\end{table}

The cube prediction target is uniform. The $gI$ selector (Section~\ref{sec:compute}) is a secondary comparator in this response study. The I-U bridge uses 1,024 starts, with its support-to-run ratio following \citet{ubridge}. Direct cube I-optimal construction uses 2,000 starts.

The responses are separate linear and dense full-quadratic truths, which are correctly specified for the working model, and four nonlinear test functions used by \citet{ubridge}: the wing-weight, OTL-circuit and borehole simulators \citep{SurjanovicBingham} and a synthetic sine function (Supplementary Table~S1). Noise has standard deviation $\phi$ times the regional standard deviation of the noiseless truth, at seven levels $\phi=0.003,0.01,0.03,0.06,0.08,0.10,0.125$.

Each design is evaluated under two fits: the quadratic working model by ordinary least squares (OLS), and a GP. OLS risk has two parts: error from approximating the response by a quadratic surface, and variance from noisy observations. We evaluate the first by numerical integration of the squared error of the surface fitted to noise-free responses. The second is $\sigma^2 I$, obtained from the target moments without simulating noise. For the correctly specified quadratic truth, only the variance term remains. The GP has a constant mean, a Gaussian correlation with a separate length scale for each factor and an estimated nugget, and is fitted by multistart profile maximum likelihood in Python. Likelihood checks identified suboptimal solutions for some training sets fitted in JMP, motivating the Python implementation. For each pool, truth and noise level, we generate four noisy response replicates, holding each selected design fixed and applying each noise vector across procedures. Prediction loss is mean-response MSE over 4,096 fixed Sobol sites. Supplementary Section~S1 describes the statistical model, and the code/data package documents fitting settings and numerical checks.

For each method, setting, truth and noise level, we average MSE over pools and replicates before taking the square root. We divide each method's root risk by that of the comparator, either best geometry or a single draw. We summarize these ratios by geometric means, giving equal weight to truths and noise levels within each setting and then to the six settings. Within a setting, correctly specified OLS contrasts are square roots of ratios of mean I and do not vary with noise level. Supplementary Table~S9 reports those variance ratios for both filters and direct A/I. Figure~\ref{fig:risksummary} also shows nonlinear truths fitted by OLS. The linear-response example is in Supplementary Section~S1. Response-study intervals are pointwise 95\% bootstrap intervals over paired blocks of two consecutive pools, conditional on the studied truths and evaluation sites and not adjusted for multiple comparisons. The code/data package gives the resampling details.

\subsection{Overall prediction and geometric cost}\label{sec:simresults}

\begin{table}[!htbp]
\centering\small
\setlength{\tabcolsep}{4pt}
\begin{tabular}{@{}lcccr@{}}
\toprule
Method & Quadratic OLS & Quadratic GP & Nonlinear GP & \shortstack{MaxPro cost\\vs best geometry (\%)} \\
\midrule
\multicolumn{5}{@{}l}{\emph{Relative to best geometry}} \\
Single FFF draw & 1.038 [1.036, 1.040] & 1.064 [1.053, 1.075] & 1.045 [1.041, 1.048] & --- \\
Raw-$S_{-1}$ filter & 0.957 [0.955, 0.958] & 0.971 [0.963, 0.979] & 0.987 [0.985, 0.990] & 1.20 \\
Target filter & 0.967 [0.965, 0.969] & 0.969 [0.961, 0.977] & 0.986 [0.984, 0.989] & 0.61 \\
$g\,I$ selector & 0.944 [0.943, 0.946] & 0.960 [0.952, 0.968] & 0.980 [0.977, 0.982] & 2.01 \\
I-optimal & 0.770 [0.769, 0.772] & 0.937 [0.927, 0.948] & 1.568 [1.551, 1.584] & undefined \\
I-U bridge & 0.827 [0.826, 0.829] & 1.079 [1.065, 1.092] & 1.230 [1.224, 1.236] & $749^{\dagger}$ \\
\bottomrule
\end{tabular}

\caption{Overall root-risk ratios in the cube response study with pointwise 95\% paired two-pool-block intervals. MSE is averaged within each setting, truth and noise level before taking roots. Log ratios weight truth/noise cells and settings equally. Values below one favor the row. MaxPro cost is the median within-pool increase over best geometry; --- denotes the ordinary candidate average, undefined denotes tied coordinates. $\dagger$ I-U bridge runs can be close in projection, substantially increasing MaxPro.}
\label{tab:responseresults}
\end{table}

Target filtering improved average prediction in all three reported families: the quadratic truth under OLS and GP fitting, and the nonlinear truths under GP fitting (Table~\ref{tab:responseresults}). Relative to best geometry, root risk fell by 3.3\%, 3.1\% and 1.4\%, respectively, at a median MaxPro increase of 0.61\%. All three intervals lie below one. Relative to a single FFF draw, the reductions were 6.8\%, 8.9\% and 5.6\% (Supplementary Table~S3). Raw filtering also improved all three averages, at a median MaxPro increase of 1.20\%.

The raw and target filters selected the same design in 1,369 of the 3,000 pools. Raw filtering gave a slightly larger reduction in quadratic-OLS root risk. For the quadratic and nonlinear GP families, paired target/raw intervals included one (Supplementary Section~S1). The $gI$ selector improved the average-risk point estimates further in all three families, at a median MaxPro cost of 2.01\%, compared with 0.61\% for target filtering. In the reference cohort (Table~\ref{tab:studymap}), either filter could select a candidate dominated on geometry and I (Supplementary Table~S4). For the target filter, a dominating candidate must have larger $D_\nu$ to offset its lower $gI_\nu$ in the score. The shape penalty therefore does not guarantee Pareto efficiency.

\subsection{Model-based designs and noise}\label{sec:heterogeneity}

\begin{figure}[!htbp]
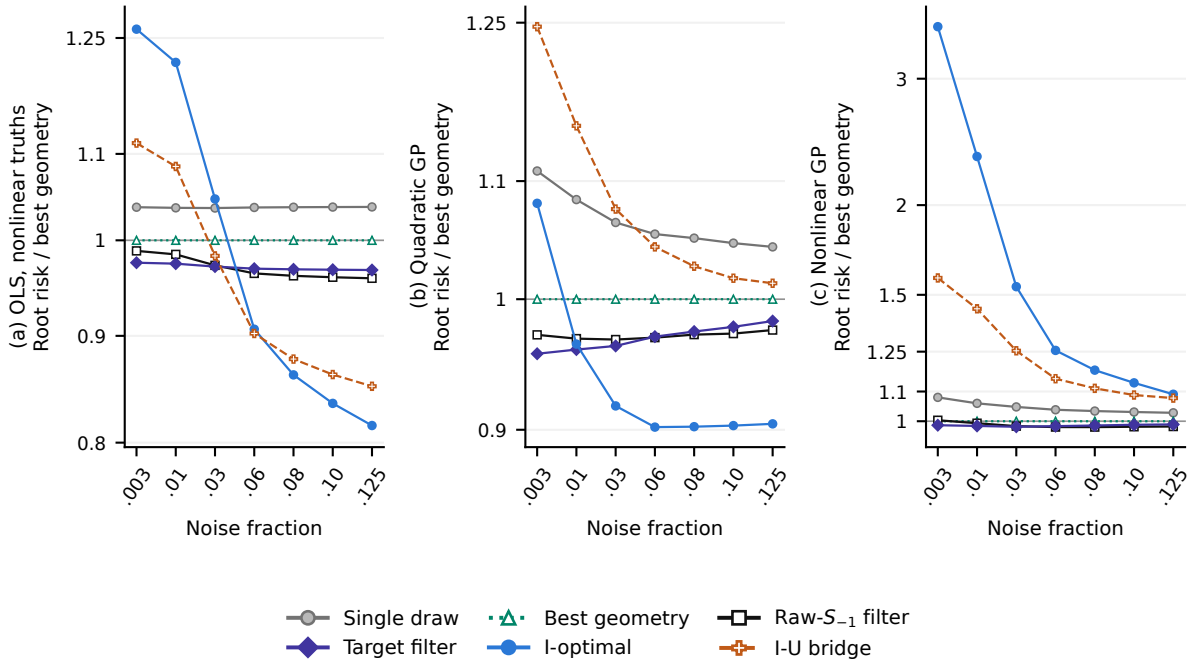

\centering
\incfig[width=0.98\linewidth]{confirmation/pool_response_summary_main.pdf}
\caption{Root-risk ratios in the cube response study relative to best geometry: (a) quadratic OLS on four nonlinear truths, (b) GP on the quadratic truth and (c) GP on nonlinear truths. Log ratios weight the indicated truths and six settings equally. Bars in (a) are pointwise 95\% paired two-pool-block intervals. Values below one favor the method; vertical scales differ.}
\label{fig:risksummary}
\end{figure}

The model-based designs make a different trade. The I-optimal design and I-U bridge reduced quadratic-OLS root risk by 23\% and 17\% relative to best geometry but increased nonlinear-GP root risk by 57\% and 23\%. Figure~\ref{fig:risksummary} shows how this depends on noise. For the nonlinear truths fitted by quadratic OLS, the I-optimal design is worse than best geometry at low noise and better from $\phi=0.06$, where its variance reduction outweighs its larger approximation error. Under GP fitting its penalty for the nonlinear truths shrinks as noise grows but does not disappear. A design that is best for the working model can predict a nonlinear response poorly, and the I-criterion, which measures working-model variance, cannot detect that on its own. The A-optimal versions gave similar working-model precision on these cubes (Supplementary Table~S9). Their response comparisons are in Table~S3.

Average gains conceal losses for some responses and noise levels (Supplementary Figure~S1). Among 252 GP cells per selector, pointwise intervals favored best geometry in 12 for target filtering, 16 for raw filtering and 16 for $gI$. Target filtering's adverse cells included ten linear-truth cells. Its overall linear-GP risk did not differ significantly from best geometry and was significantly higher than raw filtering's. The largest cell risk ratios were 1.028, 1.117 and 1.222, respectively. Raw and $gI$ losses were concentrated in the smallest two-factor setting at low noise. Target filtering also increased correctly specified OLS root risk by 0.3\% at $(2,20)$.

\subsection{Worst-decile working-model precision}\label{sec:tails}\label{sec:tailstudy}

Average prediction can conceal occasional designs with poor working-model support. The design-tail analysis examines this risk in the common six cube settings and six bounded-mixture settings (Table~\ref{tab:studymap}). Best geometry minimizes MaxPro on cubes and ACD in mixtures. For $q$ components and $n$ runs, the mixtures use $(q,n)=(3,10),(3,20),(4,15),(4,30),(5,20),(5,30)$, with one component restricted to 1--3\% and the others completing the simplex. Their quadratic Scheff\'e working model has a uniform prediction target over the feasible region (Section~\ref{sec:bounded}).

For each setting, write $I=I_\nu$ and $Z=\sqrt{I/m}$, where $m$ is the median I across all candidates in that setting's reference cohort: 50 pools of 500 candidates per cube setting and 30 pools of 500 per bounded-mixture setting (Table~\ref{tab:studymap}). The regional root-mean-square prediction standard error is $\sigma\sqrt{I}$, and $Z=1$ corresponds to the reference median I. We report each procedure's worst-decile mean $Z$ and 90th percentile, comparing both filters with a single draw and best geometry. The 96 comparisons have approximate simultaneous 95\% intervals from a paired bootstrap over blocks of pools (Supplementary Section~S2). These are tails of working-model prediction uncertainty across designs, rather than tails of fitted-response errors.

\begin{figure}[!htbp]
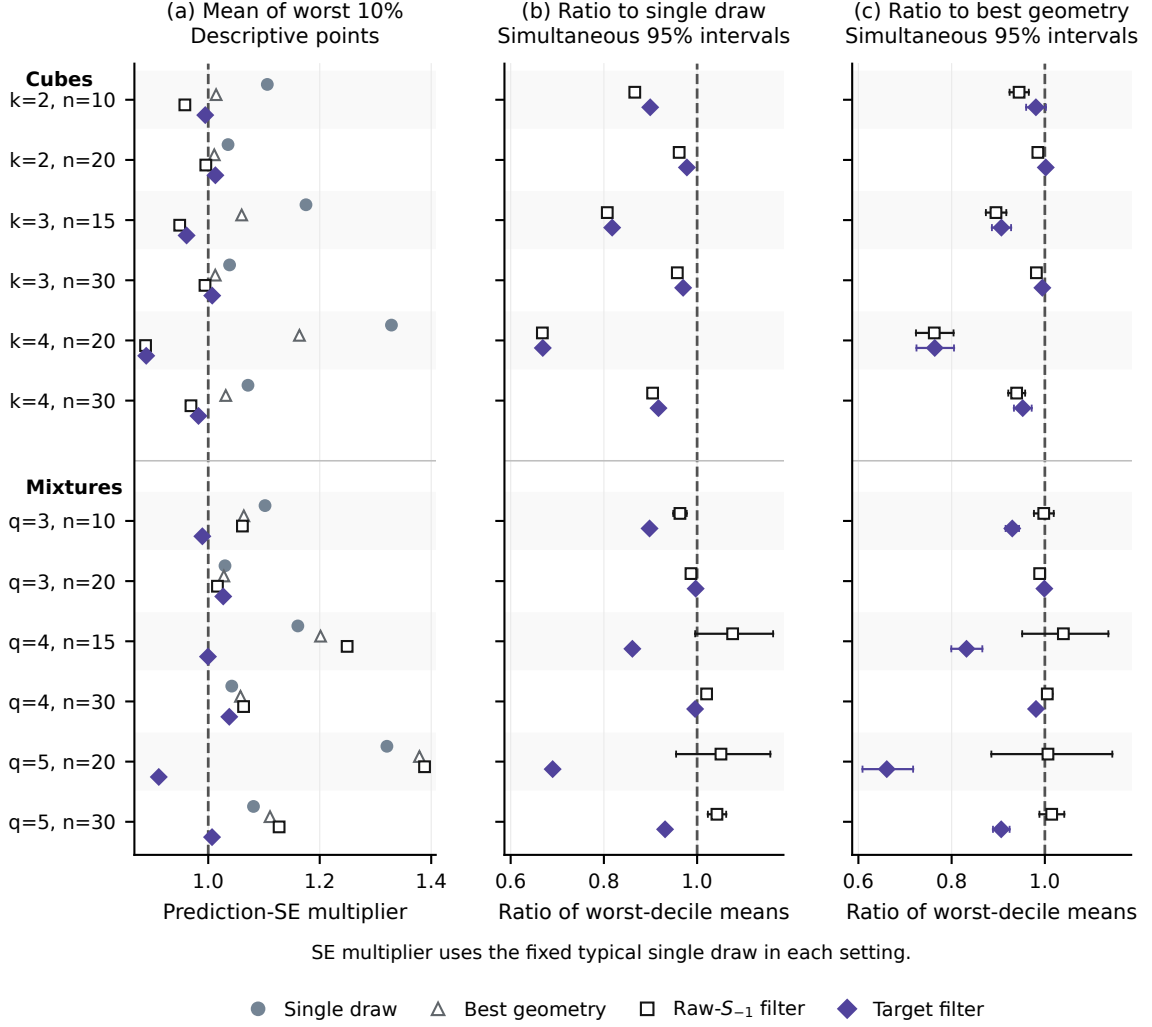

\centering
\incfig[width=0.98\linewidth]{confirmation/delivered_design_tails.pdf}
\caption{Worst-decile working-model prediction standard error in the design-tail study. (a) Mean of $Z=\sqrt{I/m}$ over the worst 10\% of designs delivered by each procedure; one corresponds to the reference median I. (b)--(c) Ratios of each filter's worst-decile mean to that of a single draw and best geometry, respectively, with approximate simultaneous 95\% intervals over the 96 planned comparisons. An interval wholly below or above one favors the filter or comparator. Across both tail summaries, 66 comparisons favor the filter, six favor the comparator and 24 show no significant difference. Each procedure supplies 500 designs per setting; the single-draw distribution uses all 250,000 candidates.}

\label{fig:centertail}
\end{figure}

Target filtering sharply reduced the worst-decile mean in cube $(k,n)=(4,20)$, where 20 runs support 15 quadratic-model coefficients. Its worst-decile mean was 0.889, about a third below a single draw and a quarter below best geometry (Figure~\ref{fig:centertail}). A fifth of single draws had prediction standard errors at least 10\% above the reference-median scale. None of the 500 designs delivered by target filtering crossed that threshold.

The scale--shape factorization describes the working-model weakness near saturation. In the reference cohort, shape accounts for about four-fifths of the worst-decile excess in log I at cube $(4,20)$ and bounded mixture $(5,20)$. At the less saturated cubes $(2,20)$ and $(3,30)$, the excess is much smaller and mostly determinant scale (Supplementary Section~S2).

Both filters improved both tail summaries over a single draw in every cube setting. Against best geometry, the raw filter improved throughout. The target filter improved in three settings and did not differ significantly in $(2,10)$, $(2,20)$ and $(3,30)$, where it often retained the best-geometry design. Of the target filter's 48 contrasts, 36 favored filtering and 12 showed no significant difference, all assessed within the original 96-comparison family. Only raw filtering in four- and five-component mixtures had significantly worse tail performance than a comparator (Section~\ref{sec:mixtureapplication}).

In an additional descriptive comparison, the $gI$ selector had lower worst-decile mean prediction standard errors than both filters in all twelve cube and bounded-mixture settings, at higher median geometry costs (Supplementary Table~S6).

\subsection{Scope and limits of the protection}\label{sec:additionaldesigns}\label{sec:additionalgp}

The same near-saturation pattern extended beyond FFF and the smaller cubes (Supplementary Table~S5). Scrambled Sobol sequences, random Latin hypercubes and Latin hypercubes optimized for centered discrepancy benefited from filtering at a larger geometric cost than FFF. At $(4,20)$, target filtering reduced the worst-decile mean by 33--44\% relative to best geometry, with median MaxPro increases of 5--8\%. Gains in the larger FFF settings were also concentrated near saturation (Supplementary Section~S2).

\section{Constrained mixtures and the choice of coordinates}\label{sec:mixtureapplication}

\subsection{Bounded mixtures}\label{sec:bounded}

The bounded mixtures of Section~\ref{sec:tailstudy} represent a formulation with one narrow-range component, such as PEG lipid in a lipid nanoparticle.

The working model is the no-intercept quadratic Scheff\'e model \citep{Scheffe1958},
\[
f(x)=\bigl(x_1,\ldots,x_q,\{x_i x_j:1\le i<j\le q\}\bigr)^\top,
\]
so $p=6,10,15$ for $q=3,4,5$, and the target $\nu$ is uniform on the feasible region. Candidates are native JMP MaxPro FFF designs. Geometry is ACD in the physical mixture coordinates, so best-geometry selection minimizes ACD, the target filter $\mathrm{ACD}/S_{-1}(C_B)$ and the raw filter $\mathrm{ACD}/S_{-1}$.

Target filtering improved regional precision at small geometric cost in the design-tail study's bounded mixtures. Using the reciprocal of the median within-pool I ratio as efficiency, it was 9.7\% above best geometry and 8.8\% above raw filtering, with a median ACD increase of 1.3\% over best geometry (Supplementary Table~S8). These efficiencies summarize paired I ratios across all 3,000 bounded-mixture pools. Figure~\ref{fig:coordinates} compares the arithmetic mean I of each procedure within each setting, using 500 pools per mean. With four or five components, raw filtering changed mean I little relative to best geometry.

The tail results reinforce this distinction (Figure~\ref{fig:centertail}). Target filtering improved both summaries over best geometry in five of the six bounded settings and over a single draw in four, with no significant difference in the remaining comparisons. In $(5,20)$ its worst-decile mean was 0.91, against 1.32 for a single draw. With four or five components, best-geometry selection was descriptively worse than a single draw: its tail summaries were 1.5--5.1\% higher. Raw filtering was significantly inferior to the comparator in six comparisons. Candidate-budget effects differed: at cube $(4,20)$, target's worst-decile means were 0.923, 0.889 and 0.883 with 50, 500 and 1,000 candidates, whereas raw's bounded $(5,20)$ mean worsened from 1.31 to 1.44 with 100 versus 1,000 candidates (Supplementary Figure~S2). A larger candidate budget can improve the raw filter's score while worsening regional prediction precision, because raw filtering balances coefficient uncertainty rather than prediction uncertainty over the region.

\subsection{LNP formulation application}\label{sec:jovelnp}

We next apply the filter to the LNP workflow of \citet{KarlEtAlJoVE2023}, which adds categorical and process factors together with a hypothetical benchmark formulation run twice. The benchmark formulation was chosen as a plausible example because the experimental recipes were proprietary. The four lipid fractions are PEG in $[0.01,0.05]$ and helper, ionizable, and cholesterol lipids in $[0.10,0.60]$, subject to their sum being one. The other factors are lipid type (H101--H103), N:P ratio, and flow rate in $[1,3]$. The published 23-run table combines a 21-run JMP FFF core with two identical H101 benchmark rows. Its FFF core was selected manually from a small pool using the I-criterion for a working model, rather than taken as an unscreened draw. This is another example of comparing complete FFF candidates, as in \citet{PiepelEtAl2019}, who used geometric criteria. Our candidates use JMP's MaxPro FFF option and keep this run structure. Supplementary Section~S4 gives the run-level rounding rules and the region over which prediction variance is averaged.

For this design screen, let $x_1,\ldots,x_4$ denote the PEG, helper, ionizable, and cholesterol fractions, in that order, and use the 14-term working-model basis
\begin{equation}\label{eq:jovelnpmodel}
\begin{aligned}
f^\top=\bigl(&x_1,x_2,x_3,x_4,x_1x_2,x_1x_3,x_1x_4,x_2x_3,\\
             &x_2x_4,x_3x_4,z_1,z_2,
               (\mathrm{N{:}P}-10)/4,\mathrm{flow}-2\bigr),
\end{aligned}
\end{equation}
where the first ten entries are the four linear and six pairwise quadratic Scheff\'e mixture terms. No extra intercept is needed because the four linear mixture terms sum to one. The lipid-type contrasts $z_1,z_2$ are centered, orthogonal and unit-variance under equal category weighting; Supplementary Section~S4 gives their values. The published response analysis starts from a richer candidate-effect set and uses SVEM to select response-specific reduced models \citep{KarlEtAlJoVE2023,KarlSVEMnet2026}.

The joint target $B$ uses uniform probability on the bounded mixture region, equal weight on the three lipid types, and independent uniform probability over the two process-factor ranges. Geometry is equal-category, stratified ACD on unique support after scaling by the factor ranges. Information uses all 23 rows.

\begin{figure}[!htbp]
\centering
\incfig[width=0.98\linewidth]{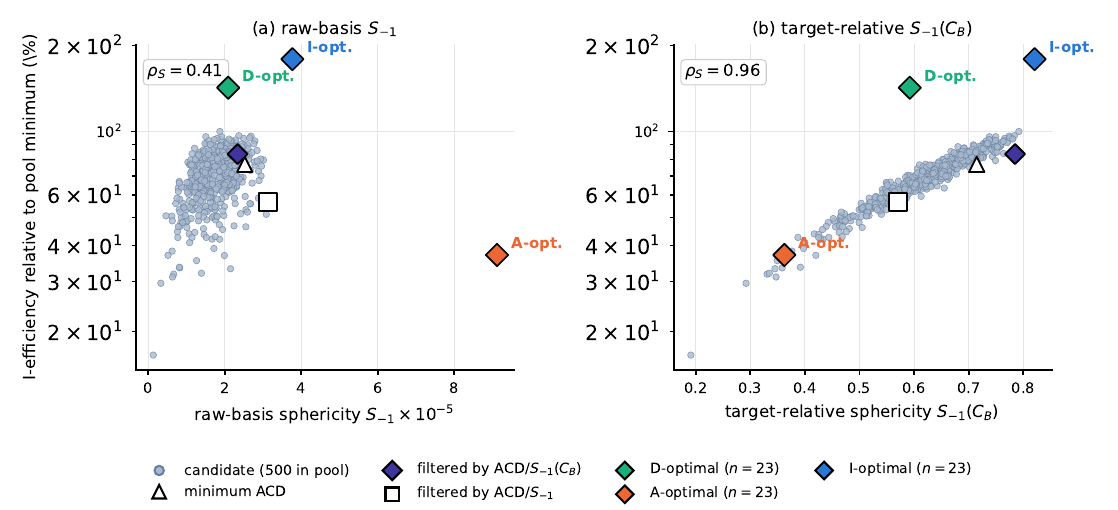}
\caption{Sphericity and I-efficiency relative to the pool minimum for the 500 candidates in the first LNP pool. For candidate $d$, the vertical quantity is $100\min_{d'\in\mathrm{pool}} I_\nu(d')/I_\nu(d)$, so larger values are favorable. Panel (a) uses raw-basis $S_{-1}$; panel (b) uses target-relative $S_{-1}(C_B)$. The marked ACD and filtering selections come from this pool; the candidates retain two fixed benchmark rows, whereas the A-, D-, and I-optimal designs were constructed with all 23 runs free and were not screening competitors.}
\label{fig:candidatesphericity}
\end{figure}

Figure~\ref{fig:candidatesphericity} shows all candidates from the first LNP pool in the reference cohort, together with A-, D-, and I-optimal reference designs. Among the 500 candidates, the Spearman correlation with I-efficiency is 0.41 for raw $S_{-1}$ and 0.96 for $S_{-1}(C_B)$. The stronger association is consistent with the prediction target built into $S_{-1}(C_B)$ through \eqref{eq:targetfactor}. Section~\ref{sec:mixtureA} explains the position of the A-optimal reference.

The 500 fresh LNP pools showed the practical difference between the two coordinate choices. Using the same median within-pool efficiency summary as above, target filtering improved by 27.5\% over minimum ACD at a median ACD increase of 2.3\%, while raw filtering lost 5.4\% in efficiency. The tails showed the same split: the target filter's worst-decile mean was 21\% below minimum ACD and the raw filter's was 12\% above it (Supplementary Section~S2). Descriptively, $gI$ lowered mean I further: relative to minimum ACD, its ratio was 0.721 versus target's 0.763, at median ACD costs of 4.0\% versus 2.3\%. Supplementary Table~S9 compares both filters with matched model-based references constructed in these same 500 fresh pools.

\subsection{Why raw filtering and raw A-optimality can lose regional precision}\label{sec:mixtureA}

The mixture results expose a mismatch between precise coefficient estimation and precise prediction. The A-optimal reference in Figure~\ref{fig:candidatesphericity} is the most spherical in Scheff\'e coordinates but among the worst for prediction inside the region. The case for A-optimal screening made by \citet{JonesAllenMoyerGoos2020} rests on precise estimates of meaningful effects. For mixtures, those effects depend on the coefficient interpretation \citep{Piepel2006} and scaling \citep{DraperPukelsheim1998}. \citet{KarlWisnowskiRushing2022} raised the resulting practical difficulty in mixture design selection.

Raw $A=\tr(C^{-1})$ weights the named coefficient variances equally. By contrast, $I_\nu=\tr(C^{-1}B)$ weights precision for regional prediction. For the mixture part of the model, a linear Scheff\'e coefficient represents a pure-component response \citep{Piepel2006}. An interaction coefficient is four times the binary-midpoint departure from the average of the two pure-component responses. Pure PEG and 50:50 PEG blends are far outside the LNP region, where PEG is restricted to 1--5\%. In the all-runs-free I-optimal reference, the PEG coefficient variance is $1.10\times10^6$, versus 10.5 for helper lipid, both omitting the common $\sigma^2$ factor. Thus a design optimized for prediction inside the region can have enormous uncertainty about the pure-PEG response. More than 99.9\% of raw A for each A-, D- and I-optimal reference comes from the four PEG-related coefficients. Raw filtering includes this same A in its score $gAD/p$. Minimizing A concentrates the design effort on these extrapolated quantities, with little weight on precision inside the region.

A- and D-optimal designs generated in JMP illustrate coding sensitivity. Both use 11 runs on the three-component full simplex, with physical proportions and the quadratic Scheff\'e model. Replacing each interaction $x_i x_j$ by its unit-peak form $4x_i x_j$ divides the associated coefficient variances by sixteen and reverses the raw-A preference (Table~\ref{tab:coding}), without changing the runs or fitted response space.

\begin{table}[!htbp]
\centering\small
\begin{tabular}{lrr}
\toprule
Ratio (A-search/D-search) & $c=1$ & $c=4$ \\
\midrule
Raw $A$ & 0.9190 & 1.0560 \\
Determinant scale $D$ & 0.9985 & 0.9985 \\
Regional $I_\nu$ & 0.8248 & 0.8248 \\
Correlation determinant, $1/p$ power & 0.9250 & 0.9250 \\
\bottomrule
\end{tabular}

\caption{Fixed-design comparison on the full simplex ($q=3$, $n=11$): A-search/D-search ratios. Here $c$ multiplies the interaction columns and $\nu$ is uniform on the full simplex. A ratio below one favors the A-search design for raw A; its preference reverses between columns. The other three ratios are unchanged. All runs have unit weight; physical proportions and pseudocomponents coincide here.}
\label{tab:coding}
\end{table}

This positive diagonal recoding preserves the determinant-scale ratio and coefficient-estimate correlations, so the raw-sphericity ratio changes through marginal-variance balance \citep[Sec.~3.3]{KarlSphericity2026}. It changes coefficient weights \citep{DraperPukelsheim1998} without necessarily matching the feasible region. Lower-bound pseudocomponent recoding can yield vertices that violate upper bounds \citep{Crosier1984}. Supplementary Section~S4 gives the LNP example. We have not evaluated the regional prediction precision of an A-optimal design constructed in those pseudocomponents. Even on the full simplex, where pure components and binary midpoints are feasible, raw A and regional I differ.

The factorization in \eqref{eq:targetfactor} shows where the A-optimal design loses. Relative to the I-optimal LNP reference in Figure~\ref{fig:candidatesphericity}, it has 2.4 times the raw sphericity (better balance), but less than half the target-relative determinant scale $D_\nu$ and only 44\% of the target-relative sphericity $S_{-1}(C_B)$. These two shortfalls multiply to give it 4.8 times the $I_\nu$. The D-optimal reference has 26\% more $I_\nu$ than the I-optimal design.

\begin{figure}[!htbp]
\centering
\incfig[width=0.98\linewidth]{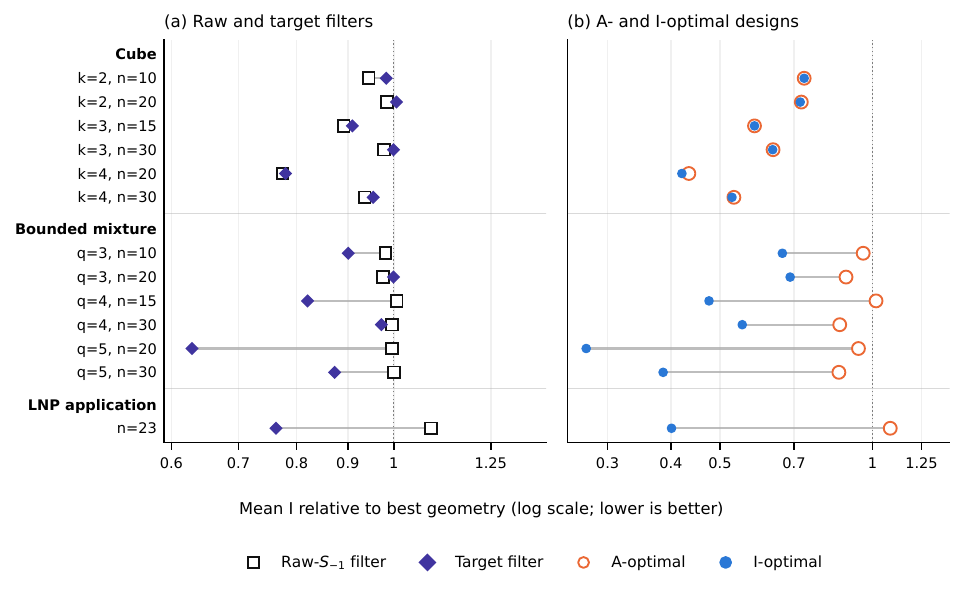}
\caption{Mean working-model prediction variance relative to best geometry, using the same 500 pools per setting in both panels (log scale; lower is better). (a) Raw and target filters. (b) Direct A- and I-optimal references; LNP references retain the benchmark pair. Points are ratios of arithmetic setting means, and gray segments connect methods rather than showing uncertainty. Supplementary Table~S9 gives all four methods with pointwise 95\% paired two-pool-block intervals.}

\label{fig:coordinates}
\end{figure}

The raw-versus-target comparison in Figure~\ref{fig:coordinates}(a) mirrors the A-versus-I comparison in panel (b). Across 500 constructions per setting, differences within each pair were small on cubes, while prediction-targeted choices generally had lower mean regional prediction variance in constrained mixtures. Across the seven constrained settings, direct I reduced mean $I_\nu$ by 31--73\% relative to best geometry. Direct A ranged from 14\% lower to 8\% higher, yet its mean $I_\nu$ was 1.3--3.4 times that of direct I. For bounded $(5,20)$, direct A improved on geometry by 6.2\% while direct I improved by 72.8\%. Direct A remained worse than geometry in LNP. Its higher mean I in bounded $(4,15)$ did not differ significantly from geometry. The A/I gap widened with component count and was larger at the smaller run size for each component count (Supplementary Table~S9).

For these regions, raw A is a poor default unless the Scheff\'e coefficients themselves, in the stated units, are the targets of the study. When collinearity makes coefficient estimates unstable in highly constrained mixture regions, \citet[p.~489]{Cornell2002} advises using the fitted model to examine response traces and contours rather than interpreting those estimates. When prediction over the region is the aim, the matched criterion is $I_\nu$ \citep{GoosJonesSyafitri2016}. Target filtering improves support for that goal within a space-filling pool, but does not reach I-optimal efficiency: the I-optimal reference in Figure~\ref{fig:candidatesphericity} has lower $I_\nu$ than the filtered choice and higher ACD than every FFF candidate in its pool. Other potential objectives include component slopes at a reference mixture for linear models \citep{Piepel2006} and weighted A-efficiency across candidate mixture models \citep{LimmunBorkowskiChomtee2018}. These were not evaluated here.

\section{Discussion and recommendations}\label{sec:discussion}

When FFF is chosen and plausible low-order effects deserve protection, target filtering is a reasonable default. It retains a complete FFF candidate and gave better average regional working-model precision at small geometric cost in these studies. Raw filtering is a practical alternative in the tested cube coding when target moments are inconvenient. The $gI$ selector puts more weight on the full prediction criterion within the same pool, with larger gains and larger geometric costs in our comparisons. The selectors were not compared at equal geometric cost.

Bridge and U-bridge designs also combine space filling with working-model precision \citep{bridge,ubridge}. I-U bridge optimizes I over a larger FFF support, whereas filtering selects an intact FFF design. Filtering suits applications where retaining space-filling geometry is the priority and working-model protection is a secondary aim. Both approaches allow flexible response analysis. In our cube study, the tested I-U bridge gained more working-model precision at far greater MaxPro cost, but had higher GP risk than best geometry for both quadratic and nonlinear truths (Table~\ref{tab:responseresults}).

When the true mean is plausibly contained in the working model, possibly using only a subset of its terms, a full-model OLS fit has no approximation bias. Direct I-optimality then targets average regional prediction variance \citep{MyersMontgomeryAndersonCook2009}. The nonlinear truths in our comparisons include complexity outside the quadratic working model, so approximation bias matters for OLS fits. Direct I also incurred nonlinear-GP losses despite its working-model precision gains. With approximation error, the trade-off also depends on noise, as Figure~\ref{fig:risksummary}(a) illustrates. Model-robust criteria offer another route \citep{DuMouchelJones1994}.

The evidence depends on the working models, targets, truths and fitting policy studied. It does not test sparse response surfaces, screening power or variable selection. Screening and prediction can favor different analyses \citep{WeeseRamseyMontgomery2018ScreeningVsPrediction}, and screening comparisons can depend on selection tuning \citep{YoungEtal2024SupportRecovery}. Singular working models require analysis-specific criteria \citep{JonesMajumdar2014}. In mixtures, the demonstrated benefit is working-model prediction precision. GP and SVEM prediction, and the effect of initial-design gains on later SVEMnet or Bayesian-optimization decisions, remain open questions.

\singlespacing
\section*{Acknowledgments}
Thank you to Bradley Jones for helpful discussions and suggestions, and to David M. Steinberg for sharing simulation code for the U-bridge study \citep{ubridge}.

\section*{Data Availability Statement}
No experimental data were collected. The supplementary document is included in this preprint. The separate computational package is not included in this version.

\section*{Funding Statement}
No funding was received for this work.

\section*{Conflict of Interest Disclosure}
The author reports no conflicts of interest.

\section*{Declaration of Generative AI and AI-Assisted Technologies in the Manuscript Preparation Process}
OpenAI's ChatGPT and Codex and Anthropic's Claude were used under the author's direction for brainstorming, critical review, restructuring and rewording text, proofreading, and LaTeX formatting. They also assisted with code development and documentation, literature crosschecks, and checks of mathematical derivations and numerical results. The author evaluated and revised the tools' suggestions, verified the substantive content, and takes full responsibility for the manuscript.

\bibliographystyle{plainnat}
\bibliography{filtering_refs}

\begin{thebibliography}{40}
\providecommand{\natexlab}[1]{#1}
\providecommand{\url}[1]{\texttt{#1}}
\expandafter\ifx\csname urlstyle\endcsname\relax
  \providecommand{\doi}[1]{doi: #1}\else
  \providecommand{\doi}{doi: \begingroup \urlstyle{rm}\Url}\fi

\bibitem[Borkowski and Piepel(2009)]{BorkowskiPiepel2009}
John~J. Borkowski and Gregory~F. Piepel.
\newblock Uniform designs for highly constrained mixture experiments.
\newblock \emph{Journal of Quality Technology}, 41\penalty0 (1):\penalty0
  35--47, 2009.
\newblock \doi{10.1080/00224065.2009.11917758}.

\bibitem[Coetzer and Haines(2017)]{CoetzerHaines2017}
Roelof Coetzer and Linda~M. Haines.
\newblock The construction of {D}- and {I}-optimal designs for mixture
  experiments with linear constraints on the components.
\newblock \emph{Chemometrics and Intelligent Laboratory Systems}, 171:\penalty0
  112--124, 2017.
\newblock \doi{10.1016/j.chemolab.2017.10.007}.

\bibitem[Cornell(2002)]{Cornell2002}
John~A. Cornell.
\newblock \emph{Experiments with Mixtures: Designs, Models, and the Analysis of
  Mixture Data}.
\newblock John Wiley \& Sons, New York, 3rd edition, 2002.
\newblock ISBN 9780471393672.
\newblock \doi{10.1002/9781118204221}.

\bibitem[Crosier(1984)]{Crosier1984}
R.~B. Crosier.
\newblock Mixture experiments: Geometry and pseudocomponents.
\newblock \emph{Technometrics}, 26\penalty0 (3):\penalty0 209--216, 1984.
\newblock \doi{10.1080/00401706.1984.10487957}.

\bibitem[Draper and Pukelsheim(1998)]{DraperPukelsheim1998}
Norman~R. Draper and Friedrich Pukelsheim.
\newblock Mixture models based on homogeneous polynomials.
\newblock \emph{Journal of Statistical Planning and Inference}, 71\penalty0
  (1--2):\penalty0 303--311, 1998.
\newblock \doi{10.1016/S0378-3758(98)00012-3}.

\bibitem[DuMouchel and Jones(1994)]{DuMouchelJones1994}
William DuMouchel and Bradley Jones.
\newblock A simple {B}ayesian modification of {D}-optimal designs to reduce
  dependence on an assumed model.
\newblock \emph{Technometrics}, 36\penalty0 (1):\penalty0 37--47, 1994.
\newblock \doi{10.1080/00401706.1994.10485399}.

\bibitem[Filov{\'a} and Harman(2020)]{FilovaHarman2020}
Lenka Filov{\'a} and Radoslav Harman.
\newblock Ascent with quadratic assistance for the construction of exact
  experimental designs.
\newblock \emph{Computational Statistics}, 35\penalty0 (2):\penalty0 775--801,
  2020.
\newblock \doi{10.1007/s00180-020-00961-9}.

\bibitem[Goos et~al.(2016)Goos, Jones, and Syafitri]{GoosJonesSyafitri2016}
Peter Goos, Bradley Jones, and Utami Syafitri.
\newblock {I}-optimal design of mixture experiments.
\newblock \emph{Journal of the American Statistical Association}, 111\penalty0
  (514):\penalty0 899--911, 2016.
\newblock \doi{10.1080/01621459.2015.1136632}.

\bibitem[Hernandez et~al.(2012)Hernandez, Lucas, and
  Sanchez]{HernandezLucasSanchez2012}
Alejandro~S. Hernandez, Thomas~W. Lucas, and Paul~J. Sanchez.
\newblock Selecting random {Latin} hypercube dimensions and designs through
  estimation of maximum absolute pairwise correlation.
\newblock In \emph{Proceedings of the 2012 Winter Simulation Conference}, pages
  280--291. IEEE, 2012.
\newblock \doi{10.1109/WSC.2012.6465112}.

\bibitem[Hvarfner et~al.(2025)Hvarfner, Eriksson, Bakshy, and
  Balandat]{HvarfnerEtAl2025}
Carl Hvarfner, David Eriksson, Eytan Bakshy, and Maximilian Balandat.
\newblock Informed initialization for {B}ayesian optimization and active
  learning.
\newblock In \emph{Advances in Neural Information Processing Systems},
  volume~38, pages 68425--68452. Curran Associates, Inc., 2025.
\newblock \doi{10.52202/085713-2300}.
\newblock URL
  \url{https://proceedings.neurips.cc/paper_files/paper/2025/hash/62dda5f9daae593009bfef839e02c891-Abstract-Conference.html}.

\bibitem[{JMP Statistical Discovery
  LLC}(2025{\natexlab{a}})]{JMPDesignEfficiency2025}
{JMP Statistical Discovery LLC}.
\newblock Design efficiency, 2025{\natexlab{a}}.
\newblock JMP Documentation, Design of Experiments Guide: Evaluate Designs.
  Publication date 07/15/2025. Available at
  \url{https://www.jmp.com/support/help/en/19.0/jmp/design-efficiency.shtml}
  (accessed 2026-06-12).

\bibitem[{JMP Statistical Discovery
  LLC}(2025{\natexlab{b}})]{JMPFFFDesignDetails19}
{JMP Statistical Discovery LLC}.
\newblock Fast flexible filling design details, 2025{\natexlab{b}}.
\newblock JMP Documentation, Design of Experiments Guide: Space-Filling
  Designs. Publication date 07/15/2025. Available at
  \url{https://www.jmp.com/support/help/en/19.0/jmp/fast-flexible-filling-design-details.shtml}
  (accessed 2026-10-10).

\bibitem[Johnson et~al.(2010)Johnson, Montgomery, Jones, and
  Parker]{JohnsonEtAl2010}
Rachel~T. Johnson, Douglas~C. Montgomery, Bradley Jones, and Peter~A. Parker.
\newblock Comparing computer experiments for fitting high-order polynomial
  metamodels.
\newblock \emph{Journal of Quality Technology}, 42\penalty0 (1):\penalty0
  86--102, 2010.
\newblock \doi{10.1080/00224065.2010.11917808}.

\bibitem[Jones and Majumdar(2014)]{JonesMajumdar2014}
Bradley Jones and Dibyen Majumdar.
\newblock Optimal supersaturated designs.
\newblock \emph{Journal of the American Statistical Association}, 109\penalty0
  (508):\penalty0 1592--1600, 2014.
\newblock \doi{10.1080/01621459.2014.938810}.

\bibitem[Jones and Steinberg(2025)]{ubridge}
Bradley Jones and David~M. Steinberg.
\newblock Unrestricted bridge ({U}-bridge) designs for settings with
  non-deterministic data.
\newblock \emph{Quality and Reliability Engineering International}, 41\penalty0
  (8):\penalty0 3405--3414, 2025.
\newblock \doi{10.1002/qre.70050}.

\bibitem[Jones et~al.(2015)Jones, Silvestrini, Montgomery, and
  Steinberg]{bridge}
Bradley Jones, Rachel~T. Silvestrini, Douglas~C. Montgomery, and David~M.
  Steinberg.
\newblock Bridge designs for modeling systems with low noise.
\newblock \emph{Technometrics}, 57\penalty0 (2):\penalty0 155--163, 2015.
\newblock \doi{10.1080/00401706.2014.923788}.

\bibitem[Jones et~al.(2021)Jones, Allen-Moyer, and
  Goos]{JonesAllenMoyerGoos2020}
Bradley Jones, Katherine Allen-Moyer, and Peter Goos.
\newblock {A}-optimal versus {D}-optimal design of screening experiments.
\newblock \emph{Journal of Quality Technology}, 53\penalty0 (4):\penalty0
  369--382, 2021.
\newblock \doi{10.1080/00224065.2020.1757391}.

\bibitem[Joseph and Hung(2008)]{JosephHung2008}
V.~Roshan Joseph and Ying Hung.
\newblock Orthogonal-maximin {Latin} hypercube designs.
\newblock \emph{Statistica Sinica}, 18\penalty0 (1):\penalty0 171--186, 2008.
\newblock URL
  \url{https://www3.stat.sinica.edu.tw/statistica/oldpdf/A18n17.pdf}.

\bibitem[Joseph et~al.(2015)Joseph, Gul, and Ba]{JosephGulBa2015}
V.~Roshan Joseph, Evren Gul, and Shan Ba.
\newblock Maximum projection designs for computer experiments.
\newblock \emph{Biometrika}, 102\penalty0 (2):\penalty0 371--380, 2015.
\newblock \doi{10.1093/biomet/asv002}.

\bibitem[Karl et~al.(2022)Karl, Wisnowski, and
  Rushing]{KarlWisnowskiRushing2022}
Andrew Karl, James Wisnowski, and Heath Rushing.
\newblock {JMP Pro 17} remedies for practical struggles with mixture
  experiments.
\newblock Discovery Summit Americas 2022, presentation 2022-US-45MP-1140, 2022.
\newblock URL
  \url{https://community.jmp.com/t5/Abstracts/JMP-Pro-17-Remedies-for-Practical-Struggles-with-Mixture/ev-p/756133}.

\bibitem[Karl(2026{\natexlab{a}})]{KarlSVEMnet2026}
Andrew~T. Karl.
\newblock {SVEMnet}: An {R} package for self-validated elastic-net ensembles
  and multi-response optimization in small-sample mixture--process experiments.
\newblock \emph{Chemometrics and Intelligent Laboratory Systems}, 271:\penalty0
  105660, 2026{\natexlab{a}}.
\newblock \doi{10.1016/j.chemolab.2026.105660}.

\bibitem[Karl(2026{\natexlab{b}})]{KarlSphericity2026}
Andrew~T. Karl.
\newblock Factoring {A}-optimality into {D}-optimality and sphericity.
\newblock arXiv:2603.07252v4 [stat.ME], 2026{\natexlab{b}}.
\newblock URL \url{https://arxiv.org/abs/2603.07252v4}.

\bibitem[Karl et~al.(2023)Karl, Essex, Wisnowski, and
  Rushing]{KarlEtAlJoVE2023}
Andrew~T. Karl, Sean Essex, James Wisnowski, and Heath Rushing.
\newblock A workflow for lipid nanoparticle ({LNP}) formulation optimization
  using designed mixture-process experiments and self-validated ensemble models
  ({SVEM}).
\newblock \emph{Journal of Visualized Experiments}, page e65200, 2023.
\newblock \doi{10.3791/65200}.
\newblock Issue 198.

\bibitem[Lekivetz and Jones(2015)]{FFF}
Ryan Lekivetz and Bradley Jones.
\newblock Fast flexible space-filling designs for nonrectangular regions.
\newblock \emph{Quality and Reliability Engineering International}, 31\penalty0
  (5):\penalty0 829--837, 2015.
\newblock \doi{10.1002/qre.1640}.

\bibitem[Lekivetz and Jones(2019)]{LekivetzJones2019}
Ryan Lekivetz and Bradley Jones.
\newblock Fast flexible space-filling designs with nominal factors for
  nonrectangular regions.
\newblock \emph{Quality and Reliability Engineering International}, 35\penalty0
  (2):\penalty0 677--684, 2019.
\newblock \doi{10.1002/qre.2429}.

\bibitem[Lemkus et~al.(2021)Lemkus, Gotwalt, Ramsey, and Weese]{LemkusEtAl2021}
Trent Lemkus, Christopher Gotwalt, Philip Ramsey, and Maria~L. Weese.
\newblock Self-validated ensemble models for design of experiments.
\newblock \emph{Chemometrics and Intelligent Laboratory Systems}, 219:\penalty0
  104439, 2021.
\newblock \doi{10.1016/j.chemolab.2021.104439}.

\bibitem[Limmun et~al.(2018)Limmun, Borkowski, and
  Chomtee]{LimmunBorkowskiChomtee2018}
Wanida Limmun, John~J. Borkowski, and Boonorm Chomtee.
\newblock Weighted {A}-optimality criterion for generating robust mixture
  designs.
\newblock \emph{Computers \& Industrial Engineering}, 125:\penalty0 348--356,
  2018.
\newblock \doi{10.1016/j.cie.2018.09.002}.

\bibitem[Lucas and Parker(2023)]{LucasParker2023}
Thomas~W. Lucas and Jeffrey~D. Parker.
\newblock The variability in design-quality measures for multiple types of
  space-filling designs created by leading software packages.
\newblock In \emph{Proceedings of the 2023 Winter Simulation Conference}, pages
  516--527. IEEE, 2023.
\newblock \doi{10.1109/WSC60868.2023.10407287}.

\bibitem[Mauchly(1940)]{Mauchly1940}
John~W. Mauchly.
\newblock Significance test for sphericity of a normal $n$-variate
  distribution.
\newblock \emph{The Annals of Mathematical Statistics}, 11\penalty0
  (2):\penalty0 204--209, 1940.
\newblock \doi{10.1214/aoms/1177731915}.

\bibitem[McKay et~al.(1979)McKay, Beckman, and
  Conover]{McKayBeckmanConover1979}
Michael~D. McKay, Richard~J. Beckman, and William~J. Conover.
\newblock A comparison of three methods for selecting values of input variables
  in the analysis of output from a computer code.
\newblock \emph{Technometrics}, 21\penalty0 (2):\penalty0 239--245, 1979.
\newblock \doi{10.1080/00401706.1979.10489755}.

\bibitem[Myers et~al.(2009)Myers, Montgomery, and
  Anderson-Cook]{MyersMontgomeryAndersonCook2009}
Raymond~H. Myers, Douglas~C. Montgomery, and Christine~M. Anderson-Cook.
\newblock \emph{Response Surface Methodology: Process and Product Optimization
  Using Designed Experiments}.
\newblock Wiley, Hoboken, NJ, 3 edition, 2009.
\newblock ISBN 9780470174463.

\bibitem[Piepel(2006)]{Piepel2006}
Greg~F. Piepel.
\newblock A note comparing component-slope, {Scheff\'{e}} and {Cox}
  parameterizations of the linear mixture experiment model.
\newblock \emph{Journal of Applied Statistics}, 33\penalty0 (4):\penalty0
  397--403, 2006.
\newblock \doi{10.1080/02664760500449170}.

\bibitem[Piepel et~al.(2019)Piepel, Stanfill, Cooley, Jones, Kroll, and
  Vienna]{PiepelEtAl2019}
Greg~F. Piepel, Bryan~A. Stanfill, Scott~K. Cooley, Bradley Jones, Jared~O.
  Kroll, and John~D. Vienna.
\newblock Developing a space-filling mixture experiment design when the
  components are subject to linear and nonlinear constraints.
\newblock \emph{Quality Engineering}, 31\penalty0 (3):\penalty0 463--472, 2019.
\newblock \doi{10.1080/08982112.2018.1517887}.

\bibitem[Pukelsheim(2006)]{Pukelsheim2006}
Friedrich Pukelsheim.
\newblock \emph{Optimal Design of Experiments}, volume~50 of \emph{Classics in
  Applied Mathematics}.
\newblock Society for Industrial and Applied Mathematics, Philadelphia, PA,
  2006.
\newblock \doi{10.1137/1.9780898719109}.
\newblock Unabridged republication of the 1993 Wiley edition.

\bibitem[Scheff{\'e}(1958)]{Scheffe1958}
Henry Scheff{\'e}.
\newblock Experiments with mixtures.
\newblock \emph{Journal of the Royal Statistical Society: Series B
  (Methodological)}, 20\penalty0 (2):\penalty0 344--360, 1958.
\newblock \doi{10.1111/j.2517-6161.1958.tb00299.x}.

\bibitem[Stallrich et~al.(2023)Stallrich, Allen-Moyer, and
  Jones]{StallrichAllenMoyerJones2023}
Jonathan Stallrich, Katherine Allen-Moyer, and Bradley Jones.
\newblock {D}- and {A}-optimal screening designs.
\newblock \emph{Technometrics}, 65\penalty0 (4):\penalty0 492--501, 2023.
\newblock \doi{10.1080/00401706.2023.2183262}.

\bibitem[Surjanovic and Bingham(2013)]{SurjanovicBingham}
Sonja Surjanovic and Derek Bingham.
\newblock Virtual library of simulation experiments: Test functions and
  datasets, 2013.
\newblock Retrieved 2026-07-12 from \url{https://www.sfu.ca/~ssurjano}.

\bibitem[Weese et~al.(2018)Weese, Ramsey, and
  Montgomery]{WeeseRamseyMontgomery2018ScreeningVsPrediction}
M.~L. Weese, P.~J. Ramsey, and D.~C. Montgomery.
\newblock Analysis of definitive screening designs: Screening vs. prediction.
\newblock \emph{Applied Stochastic Models in Business and Industry},
  34\penalty0 (2):\penalty0 244--255, 2018.
\newblock \doi{10.1002/asmb.2297}.

\bibitem[Yang et~al.(2025)Yang, Lu, and Anderson-Cook]{YangLuAndersonCook2025}
Xiankui Yang, Lu~Lu, and Christine~M. Anderson-Cook.
\newblock A scalable algorithm for generating non-uniform space-filling
  designs.
\newblock \emph{Journal of Quality Technology}, 57\penalty0 (4):\penalty0
  297--315, 2025.
\newblock \doi{10.1080/00224065.2025.2513504}.

\bibitem[Young et~al.(2024)Young, Weese, Stallrich, Smucker, and
  Edwards]{YoungEtal2024SupportRecovery}
Kade Young, Maria~L. Weese, Jonathan~W. Stallrich, Byran~J. Smucker, and
  David~J. Edwards.
\newblock A graphical comparison of screening designs using support recovery
  probabilities.
\newblock \emph{Journal of Quality Technology}, 56\penalty0 (4):\penalty0
  355--368, 2024.
\newblock \doi{10.1080/00224065.2024.2356127}.

\end{thebibliography}


\begin{thebibliography}{4}
\providecommand{\natexlab}[1]{#1}
\providecommand{\url}[1]{\texttt{#1}}
\expandafter\ifx\csname urlstyle\endcsname\relax
  \providecommand{\doi}[1]{doi: #1}\else
  \providecommand{\doi}{doi: \begingroup \urlstyle{rm}\Url}\fi

\bibitem[Jones and Steinberg(2025)]{ubridge}
Bradley Jones and David~M. Steinberg.
\newblock Unrestricted bridge ({U}-bridge) designs for settings with
  non-deterministic data.
\newblock \emph{Quality and Reliability Engineering International}, 41\penalty0
  (8):\penalty0 3405--3414, 2025.
\newblock \doi{10.1002/qre.70050}.

\bibitem[Karl(2026)]{KarlSphericity2026}
Andrew~T. Karl.
\newblock Factoring {A}-optimality into {D}-optimality and sphericity.
\newblock arXiv:2603.07252v4 [stat.ME], 2026.
\newblock URL \url{https://arxiv.org/abs/2603.07252v4}.

\bibitem[Roustant et~al.(2012)Roustant, Ginsbourger, and
  Deville]{RoustantGinsbourgerDeville2012}
Olivier Roustant, David Ginsbourger, and Yves Deville.
\newblock {DiceKriging}, {DiceOptim}: Two {R} packages for the analysis of
  computer experiments by kriging-based metamodeling and optimization.
\newblock \emph{Journal of Statistical Software}, 51\penalty0 (1):\penalty0
  1--55, 2012.
\newblock \doi{10.18637/jss.v051.i01}.

\bibitem[Surjanovic and Bingham(2013)]{SurjanovicBingham}
Sonja Surjanovic and Derek Bingham.
\newblock Virtual library of simulation experiments: Test functions and
  datasets, 2013.
\newblock Retrieved 2026-07-12 from \url{https://www.sfu.ca/~ssurjano}.

\end{thebibliography}

\section*{Author Biography}

\noindent\textbf{Andrew T.\ Karl} is a statistical consultant with Karl Statistical Services LLC in Aurora, Colorado. He received his Ph.D.\ in statistics from Arizona State University in 2012. His interests include design of experiments, mixed models, and machine-learning methods for small-sample industrial experiments.

\end{document}

\maketitle
\vspace{-3em}

\noindent This document accompanies the preprint. References below to the computational package describe separate materials that are not included in this version.

\begin{center}\small
\begin{tabular}{@{}>{\raggedright\arraybackslash}p{.30\linewidth}>{\raggedright\arraybackslash}p{.28\linewidth}>{\raggedright\arraybackslash}p{.36\linewidth}@{}}
\toprule
Evidence & Pools & Displays \\
\midrule
Common cube pools & 500 per setting & Main Tables~1, 4; Figures~1--2; Figure~S1; Tables~S2--S3 \\
Working-model tails & Same 500 per cube; 500 per bounded setting & Main Figure~3; Tables~S5--S6 \\
Other generators and larger cubes & 500 per generator or setting & Table~S5 \\
Fresh LNP application & 500 & Main Figure~5; Tables~S5, S8 \\
Matched direct A/I references & 500 per cube, bounded-mixture or LNP setting & Main Figure~5; Table~S9 \\
Reference cohort & 50 per cube; 30 per bounded setting; 30 LNP & Main Figure~4; Table~S4; tail scale $m$; S2 decomposition \\
\bottomrule
\end{tabular}
\end{center}
Every pool has 500 candidates at the principal budget. Main Figure~4 uses the first LNP pool in the reference cohort, with all-runs-free reference designs.

Main Figure~1 and Table~1 illustrate one of the 500 pools of two-factor, 20-run FFF designs generated in JMP Pro 19.1.4. We searched these pools for a case where target filtering changed the selection and lowered I, and the highest-score candidate had a visibly uneven prediction surface. Target filtering and best geometry selected the same design in 306 pools. Target filtering gave lower I in 56 and higher I in 138.

\subsection*{Computing the filter in JMP}

For target filtering, evaluate every candidate using the same working model and coding. In JMP Custom Design, select \textsf{Save X Matrix} from the red-triangle menu to save a \textsf{Moments Matrix} script in the generated design table. Right-click that script and choose \textsf{Edit} to view and copy the matrix. The moments must represent the prediction measure with common normalization. \citet[Supplement~S4]{KarlSphericity2026} gives the recipe for explicit-intercept and no-intercept Scheff\'e models. Calculate $I_\nu$ and $S_{-1}(C_B)$ from those moments, or rank by $gI_\nu D$.

For the raw cube comparator, set each factor's Coding property in Evaluate Design to the fixed bounds $[-1,1]$, use the full-quadratic model, and divide reported A-efficiency by D-efficiency. \citet[Supplement~S1]{KarlSphericity2026} gives the model and coding settings.

\section{Cube response study}\label{sec:s-cube}

\subsection{Responses, fitting and summaries}

The four nonlinear responses are wing weight, OTL circuit, borehole and synthetic sines, following \citet{ubridge} and \citet{SurjanovicBingham}. Table~\ref{tab:s-testcases} lists their active-variable order and ranges. With $z_j=\pi x_j/6$, the synthetic response is
\begin{equation}\label{eq:s-sines}
h(x)=8\sin z_1+4\sin z_2+2\sin z_3+\sin z_4+6\sin z_1\sin z_2+2\sin z_1\sin z_3,
\end{equation}
omitting terms involving unavailable coordinates. The linear truth is $h_L(x)=c\sum_{j=1}^k(-1/2)^{j-1}x_j$, with $c=\{3/\sum_{j=1}^k4^{-(j-1)}\}^{1/2}$. The full-quadratic working basis $f$ uses factors coded to $[-1,1]$ and orders its terms as the intercept, unrescaled linear terms $x_j$, lexicographically ordered products $x_i x_j$ for $i<j$, then uncentered squares $x_j^2$. For the quadratic truth, set $u_j=(-1)^{j-1}/j$ for $j=1,\ldots,p-1$, $u_0=-\sum_{j=1}^{p-1}B_{0j}u_j$, and $h_Q(x)=f(x)^\top u/(u^\top Bu)^{1/2}$. Both truths have mean zero and variance one under the uniform cube target. Every nonconstant quadratic term is active. Sparse response-surface models are not tested.

\begin{table}[!htbp]
\centering\small
\begin{tabular}{llll}
\toprule
Response & First two active inputs & Added for $k=3$ & Added for $k=4$ \\
\midrule
Wing weight & $A\,[6,10]$, $N_z\,[2.5,6]$ & $t_c\,[0.08,0.18]$ & $S_w\,[150,200]$ \\
OTL circuit & $R_{b1}\,[50,150]$, $R_{b2}\,[25,70]$ & $R_f\,[0.5,3]$ & $R_{c1}\,[1.2,2.5]$ \\
Borehole & $r_w\,[0.05,0.15]$, $L\,[1120,1680]$ & $K_w\,[9855,12045]$ & $H_u\,[990,1110]$ \\
Synthetic sines & $x_1,x_2\in[-1,1]$ & $x_3\in[-1,1]$ & $x_4\in[-1,1]$ \\
\bottomrule
\end{tabular}
\caption{Test cases and active-variable order used by \citet{ubridge}. Omitted physical inputs are fixed at their range midpoints.}
\label{tab:s-testcases}
\end{table}

Noise has standard deviation $\phi$ times the regional standard deviation of the noiseless truth. For OLS, we calculate the expected squared prediction error directly from squared bias and variance, without simulating noisy responses. The variance term uses exact polynomial moments, while nonlinear truth moments in the approximation-bias term use tensor Gauss--Legendre quadrature with 16 nodes per coordinate. For a correctly specified, full-rank, unregularized OLS fit, mean-response risk is $\sigma^2 I$, independent of the true coefficients at fixed $\sigma^2$. This property does not apply to reduced or selected models, or to GP fitting.

Each GP comparison uses four independent response replicates. Within each replicate, the same standard-normal noise vector is applied in the same run order to every arm, with pairing across truths and noise levels. For ordinary FFF, the single-draw baseline, OLS averages the risks of all 500 candidates in a pool. GP uses the first accepted candidate, held fixed across the four replicates.

Cube pools, supports and direct references were generated in JMP Pro 19 (19.1.5 and 19.1.4). Python performed U-bridge exchange, GP fitting and evaluation. The GP covariance is $\tau^2 K$, where
\[
K_{ij}=\exp\{-\textstyle\sum_l\theta_l(x_{il}-x_{jl})^2\}+\eta\mathbf1(i=j).
\]
We fit the GP response models by maximum likelihood. For fixed correlation and nugget parameters, we calculate the constant mean by generalized least squares and the process variance from the residual quadratic form divided by $n$, then optimize the resulting profile likelihood \citep[Appendix~A.2]{RoustantGinsbourgerDeville2012}. The multistart search considers zero and positive nugget, inactive-coordinate limits and the white-noise model, retaining the best valid likelihood. The code/data package's \texttt{UNIFIED\_RESPONSE.md} gives the numerical settings, verification and resampling procedures.

\subsection{Additional arms and paired comparisons}

Table~\ref{tab:s-design} gives design diagnostics for every arm, and Table~\ref{tab:s-primary} collects all arms against both baselines, including the direct A-optimal design and A-U bridge, which minimize average coefficient variance in the specified coding, and the OLS-only minimum-I candidate. On the cube the A- and I-optimal versions have similar working-model precision: their setting-mean I values differ by at most 3.2\% for the direct designs (Table~\ref{tab:s-coordinates}).

\begin{table}[!htbp]
\centering\small
\begin{tabular}{lrrrrr}
\toprule
Method & $D$ (M/O) & A-eff. (M/O) & I-eff. (M/O) & $S_{-1}$ (M/O) & MaxPro (M/O) \\
\midrule
Ordinary FFF & 1.000 & 1.000 & 1.000 & 1.000 & 1.000 \\
Best geometry & 1.035 & 1.085 & 1.072 & 1.041 & 0.929 \\
Raw-$S_{-1}$ filter & 1.073 & 1.190 & 1.140 & 1.098 & 0.942 \\
Target filter & 1.041 & 1.129 & 1.107 & 1.075 & 0.939 \\
$g\,I$ selector & 1.105 & 1.218 & 1.166 & 1.091 & 0.950 \\
A-optimal & 1.981 & 2.313 & 1.765 & 1.191 & undefined \\
I-optimal & 1.966 & 2.300 & 1.766 & 1.191 & undefined \\
A-U bridge & 1.546 & 1.789 & 1.514 & 1.156 & 7.999 \\
I-U bridge & 1.534 & 1.779 & 1.521 & 1.162 & 7.904 \\
Minimum-I FFF & 1.123 & 1.241 & 1.181 & 1.094 & 0.978 \\
\bottomrule
\end{tabular}

\caption{Cube design diagnostics. Entries are medians over six settings of ratios of setting means, with M denoting the method and O ordinary FFF. D, $S_{-1}$ and MaxPro use M/O. A- and I-efficiency use O/M variance-criterion ratios. Larger D, $S_{-1}$ and efficiency ratios, and smaller MaxPro, favor the method. Ordinary FFF averages candidates. MaxPro is undefined with tied coordinates. Dense U-bridge exchange can place runs close in projection.}
\label{tab:s-design}
\end{table}

\begin{table}[!htbp]
\centering\small
\setlength{\tabcolsep}{3pt}
\begin{tabular}{@{}lccc@{}}
\toprule
Method & Quadratic OLS & Quadratic GP & Nonlinear GP \\
\midrule
\multicolumn{4}{@{}l}{\emph{Relative to best geometry}} \\
Ordinary FFF & 1.038 [1.036, 1.040] & 1.064 [1.053, 1.075] & 1.045 [1.041, 1.048] \\
Raw-$S_{-1}$ filter & 0.957 [0.955, 0.958] & 0.971 [0.963, 0.979] & 0.987 [0.985, 0.990] \\
Target filter & 0.967 [0.965, 0.969] & 0.969 [0.961, 0.977] & 0.986 [0.984, 0.989] \\
$g\,I$ selector & 0.944 [0.943, 0.946] & 0.960 [0.952, 0.968] & 0.980 [0.977, 0.982] \\
I-optimal & 0.770 [0.769, 0.772] & 0.937 [0.927, 0.948] & 1.568 [1.551, 1.584] \\
I-U bridge & 0.827 [0.826, 0.829] & 1.079 [1.065, 1.092] & 1.230 [1.224, 1.236] \\
A-optimal & 0.7733 [0.7719, 0.7746] & 0.997 [0.975, 1.015] & 1.577 [1.569, 1.585] \\
A-U bridge & 0.8289 [0.8275, 0.8304] & 1.084 [1.071, 1.098] & 1.242 [1.235, 1.248] \\
Minimum-I FFF & 0.9388 [0.9371, 0.9404] & --- & --- \\
\midrule
\multicolumn{4}{@{}l}{\emph{Relative to a single FFF draw}} \\
Best geometry & 0.964 [0.962, 0.965] & 0.940 [0.930, 0.950] & 0.957 [0.954, 0.960] \\
Raw-$S_{-1}$ filter & 0.922 [0.921, 0.922] & 0.913 [0.905, 0.921] & 0.945 [0.942, 0.948] \\
Target filter & 0.932 [0.931, 0.933] & 0.911 [0.903, 0.919] & 0.944 [0.941, 0.947] \\
$g\,I$ selector & 0.910 [0.910, 0.910] & 0.902 [0.895, 0.910] & 0.938 [0.935, 0.941] \\
I-optimal & 0.742 [0.742, 0.742] & 0.881 [0.870, 0.891] & 1.501 [1.485, 1.516] \\
I-U bridge & 0.797 [0.797, 0.797] & 1.014 [1.002, 1.026] & 1.177 [1.171, 1.183] \\
A-optimal & 0.7451 [0.7450, 0.7452] & 0.937 [0.916, 0.954] & 1.509 [1.502, 1.517] \\
A-U bridge & 0.7987 [0.7985, 0.7990] & 1.019 [1.006, 1.032] & 1.189 [1.182, 1.195] \\
Minimum-I FFF & 0.9045 [0.9043, 0.9048] & --- & --- \\
\bottomrule
\end{tabular}

\caption{Overall root-risk ratios for all cube response study arms against both baselines, with pointwise 95\% paired two-pool-block intervals, at the seven core noise levels. Values below one favor the row method. The minimum-I FFF arm is an OLS-only diagnostic.}
\label{tab:s-primary}
\end{table}

The paired target/raw root-risk comparisons for the three reported families use pointwise 95\% two-pool-block intervals, as in main Table~4. Values above one favor raw filtering.
\begin{center}\small
\begin{tabular}{lccc}
\toprule
Comparison & Quadratic OLS & Quadratic GP & Nonlinear GP \\
\midrule
Target / raw & 1.0109 [1.0102, 1.0116] & 0.9981 [0.9941, 1.0023] & 0.9992 [0.9972, 1.0010] \\
\bottomrule
\end{tabular}

\end{center}

On the linear truth, GP root-risk ratios relative to best geometry are 0.9926 [0.9872, 0.9982] for raw filtering and 1.0050 [0.9999, 1.0104] for target filtering. Target filtering does not differ significantly from best geometry. The direct target/raw ratio is 1.0125 [1.0077, 1.0173], favoring raw. Under correct specification, the linear and quadratic OLS ratios agree.

\subsection{Heterogeneity, influence and Pareto diagnostics}

A descriptive summary of the 252 GP cells per selector (six settings, six truths and seven noise levels) shows how the aggregate gains coexist with losses. Favorable and adverse cells have pointwise intervals wholly below and above one, respectively, against best geometry. The largest risk ratio is taken over all cells.
\begin{center}\small
\begin{tabular}{lrrrr}
\toprule
Selector & Favorable & Adverse & Not significant & Largest ratio \\
\midrule
Raw-$S_{-1}$ filter & 120 & 16 & 116 & 1.117 \\
Target filter & 98 & 12 & 142 & 1.028 \\
$g\,I$ selector & 183 & 16 & 53 & 1.222 \\
\bottomrule
\end{tabular}

\end{center}
Target filtering's 12 adverse cells include two nonlinear cases and ten linear-truth cells. The largest losses for raw filtering and $gI$ occur in the two-factor, ten-run setting at low noise.

\begin{figure}[!htbp]
\centering
\includegraphics[width=0.98\linewidth]{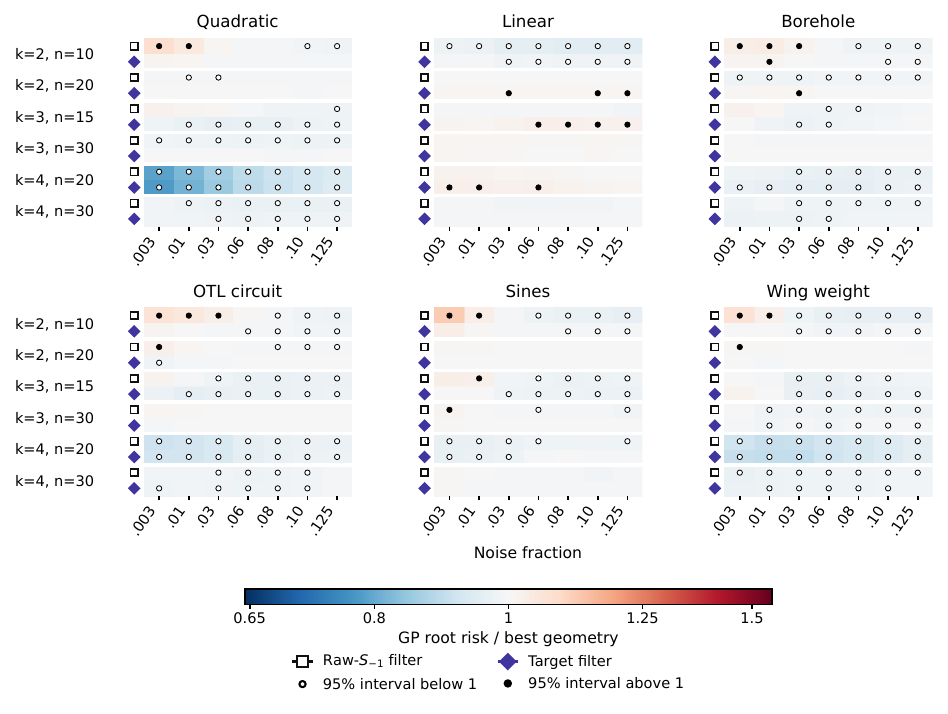}
\caption{GP root-risk ratios for both filters relative to best geometry in every truth, setting and core noise level, on a common logarithmic color scale. Within each setting, the upper row is raw filtering and the lower row target filtering. A black dot marks a pointwise 95\% paired interval entirely above one and a white dot one entirely below one. Linear denotes the correctly specified linear-truth example. All 500 pools per setting contribute. Intervals use paired two-pool blocks.}
\label{fig:gpfacets}
\end{figure}

\begin{table}[!htbp]
\centering\small
\setlength{\tabcolsep}{4pt}
\begin{tabular}{@{}lrrrrrr@{}}
\toprule
& \multicolumn{2}{c}{Nondominated (\%)} & \multicolumn{2}{c}{Median available $I$ gain (\%)} & \multicolumn{2}{c}{Maximum gain (\%)} \\
\cmidrule(lr){2-3}\cmidrule(lr){4-5}\cmidrule(l){6-7}
Setting & Raw & Target & Raw & Target & Raw & Target \\
\midrule
$k=2,n=10$ & 78 & 82 & 0.70 & 1.88 & 3.76 & 10.46 \\
$k=2,n=20$ & 80 & 70 & 0.47 & 2.70 & 3.88 & 6.18 \\
$k=3,n=15$ & 76 & 62 & 3.23 & 4.86 & 6.86 & 14.15 \\
$k=3,n=30$ & 76 & 68 & 1.43 & 3.41 & 4.60 & 11.85 \\
$k=4,n=20$ & 74 & 70 & 4.13 & 4.13 & 10.41 & 11.91 \\
$k=4,n=30$ & 76 & 58 & 2.09 & 3.10 & 7.79 & 7.79 \\
\bottomrule
\end{tabular}

\caption{Pareto diagnostics from the reference cohort's 50 cube pools per setting. A candidate dominates a selection if its MaxPro and I are no larger and at least one is smaller. The available I gain is $100(1-I_{\min}/I_{\rm selected})$, where $I_{\min}$ is the least I among candidates with no greater MaxPro. Gain medians and maxima use only dominated selections. All-pool medians are zero.}
\label{tab:s-pareto}
\end{table}

Raw and target selections are nondominated in 58--82\% of these cube pools (Table~\ref{tab:s-pareto}). Minimum MaxPro, minimum I and $gI$ selections are nondominated in every observed pool. A positive-product $gI$ minimum has this property by construction. Endpoint minima require an appropriate tie-break to guarantee it. The available gains describe possible improvements within each pool.

\section{Design-tail study}\label{sec:s-tails}

\subsection{FFF pools in cubes and bounded mixtures}

The design-tail study has 500 pools of 500 MaxPro FFF candidates in each of six cube and six bounded-mixture settings, generated in JMP Pro 19.1.4. Each mixture pool has its own ACD reference sample. In $Z=\sqrt{I/m}$, $m$ is the median candidate I in the setting's separate reference cohort: 50 pools for each cube setting and 30 for each bounded-mixture setting. For a single draw, the summaries use all 250,000 candidates in the setting's 500 study pools.

The 96 comparisons have approximate simultaneous 95\% intervals, calibrated by the bootstrap distribution of the maximum standardized deviation across the log ratios. Two consecutive 500-candidate pools share a generation unit, so paired bootstrap intervals resample those pairs together, retaining all selectors and ordinary candidates within each pair. For candidate-budget comparisons, resampling also preserves the nested partitions and their ACD reference samples. Intervals wholly below or above one favor the filter or comparator, respectively. Intervals containing one show no significant difference. The worst-decile mean averages the upper 10\% of the distribution by empirical mass.

The package tabulates exceedance probabilities at $Z\ge1.10$ and $Z\ge1.25$ for all twelve settings, with descriptive pointwise 95\% Clopper--Pearson intervals under the independent-generator model.

\begin{figure}[!htbp]
\centering
\includegraphics[width=.90\linewidth]{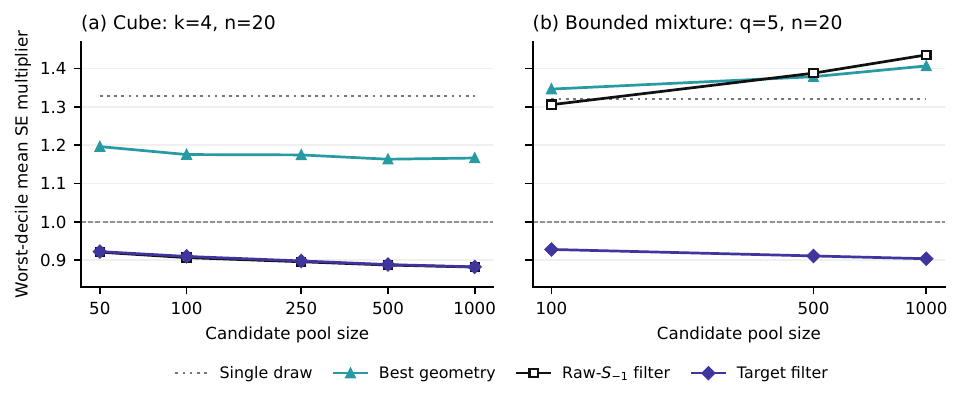}
\caption{Pool-size examples: cube $(4,20)$ and bounded mixture $(5,20)$. Curves show the worst-decile mean of $\sqrt{I/m}$ for each procedure. A value of one corresponds to the reference median I. Candidates are reused across budgets, so points are not independent experiments. Mixture subgroups use their particular ACD reference samples.}
\label{fig:s-poolsize}
\end{figure}

To describe the source of poor working-model precision, order candidates by I within each reference pool. Let $T$ average over its worst decile and $M$ average over its two middle-ranked candidates, using those same candidates for every quantity. The identity $\log I=\log p-\log D_\nu-\log S_{-1}(C_B)$ gives the additive decomposition
\[
(T-M)\log I=-(T-M)\log D_\nu-(T-M)\log S_{-1}(C_B).
\]
Average these three components over pools within a setting. The shape share is the mean shape component divided by mean log-I excess. This common reference makes the split additive, unlike centering each factor at its own marginal median. Four settings illustrate the contrast between near-saturated and larger designs:
\begin{center}\small
\begin{tabular}{@{}lrrr@{}}
\toprule
Setting & $n/p$ & Log-$I$ excess & Shape share (\%) \\
\midrule
Cube $(4,20)$ & 1.33 & 0.545 & 80 \\
Mixture $(5,20)$ & 1.33 & 0.540 & 81 \\
Cube $(2,20)$ & 3.33 & 0.070 & 8 \\
Cube $(3,30)$ & 3.00 & 0.074 & 20 \\
\bottomrule
\end{tabular}

\end{center}
The cube gains in Figure~\ref{fig:s-poolsize} plateau, whereas a larger pool can worsen raw-filter precision in constrained mixtures.

\subsection{Other generators and larger cubes}

Three additional design-only studies tested the scope of the tail results. The first used 500 pools of 500 candidates in each of the six cube settings for each of three generators: scrambled Sobol, plain Latin hypercube sampling (LHS), and LHS optimized for centered discrepancy (LHS-CD). All three used MaxPro for pool selection and the same quadratic basis, cube target and fixed median-I scales as the FFF study. Sobol candidates were non-power-of-two prefixes of scrambled nets and do not retain the full nets' balance guarantee. The second used 500 fresh LNP pools with the construction, executed-run mapping, benchmark pair and target of the paper's LNP application. The third used 500 FFF pools of 500 candidates in each larger-cube setting $(k,n)=(5,26),(5,42),(6,33),(6,56)$. Each larger-cube median-I scale came from a separate set of 5,000 designs. These studies evaluate working-model precision without fitting responses.

The endpoints are ratios of 90th percentiles and worst-decile means of working-model prediction standard error. Paired two-pool-block intervals have approximate simultaneous 95\% coverage separately over the 144 generator, eight LNP and 32 larger-cube contrasts. Two additional LNP target/raw bounds are derived conservatively from the eight jointly covered contrasts. Coverage is within each family, not across all studies. Table~\ref{tab:s-additionaltails} gives seven primary comparisons.

\begin{table}[!htbp]
\centering\small
\begin{tabular}{@{}lll@{}}
\toprule
Setting or generator & Comparison & Worst-decile ratio [interval]\\
\midrule
Sobol, $(4,20)$ & Target / best geometry & 0.594 [0.512,0.691]\\
LHS, $(4,20)$ & Target / best geometry & 0.564 [0.492,0.646]\\
LHS-CD, $(4,20)$ & Target / best geometry & 0.666 [0.608,0.728]\\
Fresh LNP & Target / minimum ACD & 0.788 [0.759,0.818]\\
Fresh LNP & Target / raw & 0.704 [0.668,0.742]\\
FFF, $(5,26)$ & Target / best geometry & 0.665 [0.625,0.707]\\
FFF, $(6,33)$ & Target / best geometry & 0.585 [0.546,0.627]\\
\bottomrule
\end{tabular}
\caption{Primary worst-decile comparisons in the additional design-only studies. Ratios below one favor the numerator. Intervals have approximate simultaneous 95\% coverage within their respective 144-, eight- and 32-contrast families. The LNP target/raw interval is derived conservatively from the eight-contrast family.}
\label{tab:s-additionaltails}
\end{table}

Filtering was significantly better in 141 of the 144 other-generator contrasts. The three comparisons showing no significant difference all involve target filtering versus best geometry at $(2,20)$: the Sobol 90th percentile and both LHS-CD summaries. Unlike FFF, these generators put no points on cube faces, so absolute cross-generator differences also reflect boundary coverage. At $(4,20)$, median target-filter MaxPro increases are 6.04\%, 7.70\% and 4.80\% for Sobol, LHS and LHS-CD.

For larger cubes $(5,42)$ and $(6,56)$, target/best-geometry worst-decile ratios are 0.939 [0.929,0.949] and 0.942 [0.933,0.951]. Median MaxPro costs in the two near-saturated settings are 4.00\% and 4.27\%.

In the fresh LNP pools, six of the ten tail ratios favor filtering. Raw filtering is significantly worse than minimum ACD for both summaries (worst-decile ratio 1.118 [1.055,1.185]) and does not differ significantly from a single draw for either summary. Table~\ref{tab:s-samepool} gives the corresponding paired median I efficiencies and pointwise intervals.

\subsection{Descriptive \texorpdfstring{$gI$}{gI} tail comparison}

\begin{table}[!htbp]
\centering\small
\setlength{\tabcolsep}{4pt}
\begin{tabular}{@{}lrrrrrr@{}}
\toprule
& \multicolumn{3}{c}{Worst-decile ratio to best geometry}
& \multicolumn{3}{c}{Median geometry increase (\%)} \\
\cmidrule(lr){2-4}\cmidrule(l){5-7}
Setting & Raw & Target & $gI$ & Raw & Target & $gI$ \\
\midrule
$k=2,n=10$ & 0.945 & 0.981 & 0.939 & 0.69 & 0.00 & 1.53 \\
$k=2,n=20$ & 0.985 & 1.002 & 0.975 & 0.27 & 0.00 & 0.86 \\
$k=3,n=15$ & 0.895 & 0.907 & 0.868 & 2.34 & 2.05 & 3.17 \\
$k=3,n=30$ & 0.982 & 0.995 & 0.960 & 0.58 & 0.15 & 1.56 \\
$k=4,n=20$ & 0.763 & 0.764 & 0.739 & 3.46 & 3.21 & 3.61 \\
$k=4,n=30$ & 0.939 & 0.953 & 0.915 & 1.76 & 1.35 & 2.42 \\
\addlinespace
$q=3,n=10$ & 0.998 & 0.930 & 0.881 & 2.44 & 1.61 & 2.71 \\
$q=3,n=20$ & 0.989 & 0.999 & 0.947 & 1.74 & 0.67 & 1.85 \\
$q=4,n=15$ & 1.040 & 0.832 & 0.761 & 2.79 & 1.62 & 4.50 \\
$q=4,n=30$ & 1.005 & 0.981 & 0.909 & 1.84 & 0.70 & 3.01 \\
$q=5,n=20$ & 1.007 & 0.661 & 0.613 & 2.65 & 2.64 & 4.47 \\
$q=5,n=30$ & 1.015 & 0.906 & 0.835 & 2.01 & 1.42 & 3.54 \\
\bottomrule
\end{tabular}

\caption{Descriptive working-model-tail comparisons from the saved FFF pools. Ratios compare worst-decile means of $\sqrt{I/m}$ with best geometry, with values below one favoring the selector. Costs are median within-pool percentage increases in MaxPro (cubes) or ACD (mixtures). These additional $gI$ comparisons have no simultaneous intervals and lie outside the 96-contrast family.}
\label{tab:s-selector-tails}
\end{table}

The $gI$ selector has lower estimated worst-decile means and higher median geometry costs than both filters in all twelve settings (Table~\ref{tab:s-selector-tails}). In the 500 fresh LNP pools, its mean I relative to minimum ACD was 0.721, compared with 0.763 for target filtering. Median ACD costs were 4.03\% and 2.35\%, respectively.

\section{Additional response diagnostics}\label{sec:s-gp}

The higher-noise OLS extension uses $\phi=0.25$ and $0.5$ and is analyzed separately from the seven-level core. Table~\ref{tab:s-gp} averages log root-risk ratios over the four nonlinear truths, two noise levels and six settings. Direct I and the U-bridges gain more when observation variance is large enough to outweigh approximation error. This does not remove their nonlinear-GP penalties in the core comparisons.

\begin{table}[!htbp]
\centering\small
\begin{tabular}{lc}
\toprule
Method & Nonlinear OLS, higher noise \\
\midrule
Raw-$S_{-1}$ filter & 0.957 [0.955, 0.959] \\
Target filter & 0.967 [0.966, 0.969] \\
$g\,I$ selector & 0.945 [0.944, 0.947] \\
A-optimal & 0.781 [0.780, 0.783] \\
I-optimal & 0.778 [0.777, 0.780] \\
A-U bridge & 0.833 [0.832, 0.835] \\
I-U bridge & 0.831 [0.830, 0.833] \\
\bottomrule
\end{tabular}

\caption{Nonlinear OLS root risk relative to best geometry at $\phi=0.25$ and $0.5$, with pointwise 95\% paired two-pool-block intervals.}
\label{tab:s-gp}
\end{table}

\section{Constrained mixtures and reference designs}\label{sec:s-samepool}

Table~\ref{tab:s-samepool} compares the filters within 500-candidate pools. The cube and bounded-mixture rows use the design-tail study, and the LNP rows use the fresh application study.

\begin{table}[!htbp]
\centering\small
\setlength{\tabcolsep}{3pt}
\begin{tabular}{@{}lrrr@{}}
\toprule
Filter / comparator & \shortstack{Geometry ratio\\filter/comparator} & \shortstack{I-efficiency gain\\comparator/filter} & \shortstack{I counts\\lower/same/higher} \\
\midrule
Cube raw / best geometry & 1.012 & 1.049 & 2112/691/197 \\
Bounded target / best geometry & 1.013 & 1.097 & 2417/202/381 \\
Bounded target / raw & 0.992 & 1.088 & 2268/34/698 \\
LNP target / best geometry & 1.023 & 1.275 [1.255, 1.304] & 458/40/2 \\
LNP target / raw & 0.976 & 1.344 [1.318, 1.383] & 481/3/16 \\
LNP raw / best geometry & 1.048 & 0.946 [0.923, 0.979] & 207/5/288 \\
\bottomrule
\end{tabular}

\caption{Same-pool comparisons from 3,000 cube pools, 3,000 bounded-mixture pools and 500 LNP pools. The cube row reports raw filtering. Table~\ref{tab:s-coordinates} includes both filters. Geometry is MaxPro for cubes and ACD for mixtures. Geometry entries are median filter/comparator ratios, and I-efficiency entries are $1/\operatorname{median}(I_{\rm filter}/I_{\rm comparator})$, using $I_\nu$ in constrained regions. Counts give lower/same/higher I for the filter, with lower favorable. Brackets on LNP efficiencies are pointwise 95\% paired-pool intervals. The other rows are descriptive point estimates.}
\label{tab:s-samepool}
\end{table}

\subsection*{Bounded-mixture construction and prediction target}

A trial construction with FFF's centroid option in raw proportions concentrated the 1--3\% component near the middle of its range and gave very poor working-model support. The reported study uses JMP MaxPro FFF pools. In each pool, ACD is computed in the physical mixture coordinates using a 16,384-site scrambled Sobol reference sample. The moment matrix $B$ uses the uniform prediction measure on the constrained simplex: Dirichlet$(1,\ldots,1)$ conditional on $0.01\le x_1\le0.03$. The restricted component has density proportional to $(1-t)^{q-2}$, and the remaining proportions divided by $1-t$ are conditionally uniform. Analytic polynomial moments are computed under this measure.

Range-scaled ACD was rescored on the same 500 pools per bounded-mixture setting. Relative to physical-coordinate ACD selection, its mean-I ratios were 1.076, 1.025, 0.963, 0.999, 0.883 and 0.975 in the displayed setting order. Thus scaling improved the point estimate in four of six settings and worsened it in two. Its 11.7\% improvement at $(5,20)$ was smaller than target filtering's 37.1\% improvement over the same physical-ACD baseline. Geometry scaling matters, but this comparison does not remove the distinction between coefficient and regional prediction objectives.

\subsection*{LNP coding, execution and coverage}

With coordinates ordered as H101, H102, and H103, the lipid-type contrasts are
\[
z_1=(\sqrt{3/2},-\sqrt{3/2},0),\qquad
z_2=(1/\sqrt{2},1/\sqrt{2},-2/\sqrt{2}).
\]
They are centered, orthogonal and unit-variance under equal category weighting.

For lower bounds $L_i$, lower-bound pseudocomponents are $u_i=(x_i-L_i)/(1-\sum_j L_j)$. The LNP bounds sum to 0.31, so the PEG pseudo-vertex $u_1=1$ corresponds to $x_1=0.01+(1-0.31)=0.70$, violating its 0.05 upper bound. We did not construct or compare A-optimal designs using these pseudocomponents.

For LNP execution, mixture fractions are rounded to $10^{-4}$ by bounded largest remainders preserving their sum, with ties in PEG/helper/ionizable/cholesterol order. Generated N:P $r\in[5,15]$ maps to $\min\{14,\max[6,2\lfloor r/2+1/2\rfloor]\}$. Flow $v\in[1,3]$ maps to $\min\{3,\max[1,\lfloor2v+1/2\rfloor/2]\}$. N:P therefore has execution levels 6, 8, 10, 12 and 14, and flow has levels 1, 1.5, 2, 2.5 and 3. The two appended hypothetical H101 benchmarks have proportions $(.01,.33,.33,.33)$, N:P 10 and flow 1. The continuous prediction target retains N:P in $[6,14]$ and flow in $[1,3]$. The mixture target is uniform conditional on all four lipid bounds, with moments integrated by inclusion--exclusion. Centered category/process moments give the remaining block $\operatorname{diag}(1,1,1/3,1/3)$.

Let $U_c$ be the unique executed support of lipid type $c$, and let $T$ scale the four proportions and two process factors by widths $(.04,.50,.50,.50,8,2)$. For $M=16,384$ independently scrambled reference sites per category,
\[
\mathrm{ACD}=\frac1{3M}\sum_{c=1}^3\sum_{m=1}^M\min_{z\in U_c}\|Tr_{cm}-Tz\|.
\]
Nearest neighbors must therefore share the reference site's lipid type. Category is not an extra Euclidean coordinate. Information retains both benchmark replicates.

\subsection{Matched reference designs}\label{sec:s-coordinates}

The matched direct A- and I-reference searches use 500 constructions per setting. Cube references use native JMP searches. Bounded-mixture and LNP references optimize on their feasible supports with finite search budgets. LNP references retain the same two benchmark rows as their paired FFF candidates. Table~\ref{tab:s-coordinates} compares both filters and these references using ratios of arithmetic mean regional prediction variance to best geometry, with pointwise 95\% paired two-pool-block intervals.

\begin{table}[!htbp]
\centering\small
\setlength{\tabcolsep}{4pt}
\begin{tabular}{@{}lcccc@{}}
\toprule
Setting & Raw filter & Target filter & Direct A & Direct I \\
\midrule
\multicolumn{5}{l}{\emph{Cube (500 pools per setting)}} \\
$k=2$, $n=10$ & 0.944 & 0.983 & 0.733 & 0.733 \\
 & {\scriptsize[0.939, 0.949]} & {\scriptsize[0.978, 0.987]} & {\scriptsize[0.730, 0.736]} & {\scriptsize[0.730, 0.736]} \\[2pt]
$k=2$, $n=20$ & 0.985 & 1.007 & 0.724 & 0.720 \\
 & {\scriptsize[0.983, 0.986]} & {\scriptsize[1.005, 1.008]} & {\scriptsize[0.722, 0.725]} & {\scriptsize[0.719, 0.722]} \\[2pt]
$k=3$, $n=15$ & 0.891 & 0.909 & 0.585 & 0.585 \\
 & {\scriptsize[0.883, 0.899]} & {\scriptsize[0.902, 0.917]} & {\scriptsize[0.580, 0.590]} & {\scriptsize[0.580, 0.590]} \\[2pt]
$k=3$, $n=30$ & 0.978 & 1.000 & 0.637 & 0.635 \\
 & {\scriptsize[0.976, 0.981]} & {\scriptsize[0.997, 1.002]} & {\scriptsize[0.635, 0.638]} & {\scriptsize[0.634, 0.637]} \\[2pt]
$k=4$, $n=20$ & 0.775 & 0.780 & 0.434 & 0.421 \\
 & {\scriptsize[0.760, 0.789]} & {\scriptsize[0.765, 0.794]} & {\scriptsize[0.426, 0.442]} & {\scriptsize[0.413, 0.428]} \\[2pt]
$k=4$, $n=30$ & 0.935 & 0.954 & 0.533 & 0.528 \\
 & {\scriptsize[0.930, 0.941]} & {\scriptsize[0.948, 0.960]} & {\scriptsize[0.530, 0.536]} & {\scriptsize[0.525, 0.531]} \\[2pt]
\multicolumn{5}{l}{\emph{Bounded mixture (500 pools per setting)}} \\
$q=3$, $n=10$ & 0.982 & 0.901 & 0.959 & 0.664 \\
 & {\scriptsize[0.974, 0.989]} & {\scriptsize[0.896, 0.906]} & {\scriptsize[0.942, 0.976]} & {\scriptsize[0.661, 0.667]} \\[2pt]
$q=3$, $n=20$ & 0.975 & 1.000 & 0.887 & 0.688 \\
 & {\scriptsize[0.972, 0.978]} & {\scriptsize[0.997, 1.002]} & {\scriptsize[0.879, 0.895]} & {\scriptsize[0.686, 0.689]} \\[2pt]
$q=4$, $n=15$ & 1.007 & 0.820 & 1.017 & 0.476 \\
 & {\scriptsize[0.985, 1.032]} & {\scriptsize[0.810, 0.831]} & {\scriptsize[0.983, 1.054]} & {\scriptsize[0.470, 0.481]} \\[2pt]
$q=4$, $n=30$ & 0.996 & 0.972 & 0.862 & 0.553 \\
 & {\scriptsize[0.992, 1.001]} & {\scriptsize[0.968, 0.976]} & {\scriptsize[0.854, 0.870]} & {\scriptsize[0.552, 0.555]} \\[2pt]
$q=5$, $n=20$ & 0.996 & 0.629 & 0.938 & 0.272 \\
 & {\scriptsize[0.959, 1.034]} & {\scriptsize[0.612, 0.645]} & {\scriptsize[0.899, 0.981]} & {\scriptsize[0.265, 0.279]} \\[2pt]
$q=5$, $n=30$ & 1.000 & 0.873 & 0.859 & 0.386 \\
 & {\scriptsize[0.990, 1.010]} & {\scriptsize[0.866, 0.880]} & {\scriptsize[0.845, 0.872]} & {\scriptsize[0.384, 0.389]} \\[2pt]
\multicolumn{5}{l}{\emph{LNP (500 pools per setting)}} \\
LNP, $n=23$ & 1.089 & 0.763 & 1.084 & 0.401 \\
 & {\scriptsize[1.060, 1.117]} & {\scriptsize[0.749, 0.776]} & {\scriptsize[1.060, 1.110]} & {\scriptsize[0.394, 0.408]} \\[2pt]
\bottomrule
\end{tabular}

\caption{Mean regional I relative to best geometry, with pointwise 95\% paired two-pool-block intervals, in the matched 500-pool cohorts. LNP references retain the benchmark pair.}
\label{tab:s-coordinates}
\end{table}

Direct A/direct I ranges from 1.000 to 1.032 on cubes and from 1.289 to 3.448 in bounded mixtures, and is 2.703 for LNP. Direct A improves its own raw-A criterion relative to direct I in every constrained setting while having higher regional I. This supports an objective trade-off. Direct A has significantly higher regional I than best geometry for LNP (1.084 [1.060, 1.110]), but does not differ significantly in bounded $(4,15)$ (1.017 [0.983, 1.054]). Raw filtering has lower mean I than target filtering in all six cubes and bounded $(3,20)$.

\FloatBarrier
\begingroup\small
\setlength{\bibsep}{0pt}
\bibliographystyle{plainnat}
\bibliography{filtering_refs}
\endgroup